\documentclass[aps,twocolumn,pra]{revtex4-1}
\usepackage[T1]{fontenc}
\usepackage[latin9]{inputenc}
\usepackage{color}
\usepackage{amsmath}
\usepackage{amssymb}
\usepackage{esint}
\usepackage{graphicx}
\usepackage{multirow}
\usepackage{amsfonts}
\usepackage{booktabs}
\usepackage{todonotes}
\makeatletter

\@ifundefined{textcolor}{}
{
\definecolor{BLACK}{gray}{0}
 \definecolor{WHITE}{gray}{1}
 \definecolor{RED}{rgb}{1,0,0}%
 \definecolor{GREEN}{rgb}{0,1,0}
 \definecolor{BLUE}{rgb}{0,0,1}
 \definecolor{CYAN}{cmyk}{1,0,0,0}
 \definecolor{MAGENTA}{cmyk}{0,1,0,0}
 \definecolor{YELLOW}{cmyk}{0,0,1,0}
}
\graphicspath{{}{images/}}
\newcommand{\abs}[1]{\lvert #1 \rvert}

\makeatother
\begin{document}

\title{Double stage nematic bond-ordering above double stripe magnetism: application to BaTi$_{2}$Sb$_{2}$O}

\author{G. Zhang$^{1,2}$, J.~K.~Glasbrenner$^{3}$, R. Flint$^{1,2}$,
  I.~I.~Mazin$^{4}$, R. M. Fernandes$^{5}$}

\affiliation{$^{1}$Department of Physics and Astronomy, Iowa State University,
  12 Physics Hall, Ames, Iowa 50011, USA}
\affiliation{$^{2}$Division of Materials Science and Engineering, Ames
  Laboratory, U.S.  DOE, Ames, Iowa 50011, USA}
\affiliation{$^{3}$National Research Council/Code 6393, Naval Research
  Laboratory, Washington, DC 20375, USA}
\affiliation{$^{4}$Code 6393, Naval Research Laboratory, Washington, DC 20375,
  USA}
\affiliation{$^{5}$School of Physics and Astronomy, University of Minnesota,
  Minneapolis, MN 55455, USA}

\begin{abstract}
  Spin-driven nematicity, or the breaking of the point-group symmetry of the lattice without
  long-range magnetic order, is clearly quite important in iron-based
  superconductors. From a symmetry point of view, nematic order can be described as a coherent locking of spin fluctuations in two interpenetrating Néel
  sublattices with ensuing nearest-neighbor bond order and an absence of static
  magnetism. Here, we argue that the low-temperature state of the recently
  discovered superconductor BaTi$_{2}$Sb$_{2}$O is a strong candidate for a more exotic form of spin-driven nematic order, in which
  fluctuations occurring in \textit{four} N\'{e}el sublattices promote both
  nearest- and next-nearest neighbor bond order. We develop a low-energy
  field theory of this state and show that it can have, as a function of temperature, up to two separate bond-order phase transitions -- namely, one that breaks rotation symmetry and one that breaks reflection and translation symmetries of the lattice.  The resulting state has an orthorhombic lattice distortion, an
  intra-unit-cell charge density wave, and no long-range magnetic order, all
  consistent with reported measurements of the low-temperature phase of
  BaTi$_{2}$Sb$_{2}$O. We then use density functional theory calculations to
  extract exchange parameters to confirm that the model is applicable to
  BaTi$_{2}$Sb$_{2}$O.
\end{abstract}

\maketitle

\section{Introduction}

\label{sec:introduction}

\begin{figure}[th]
\includegraphics[width=0.48\textwidth]{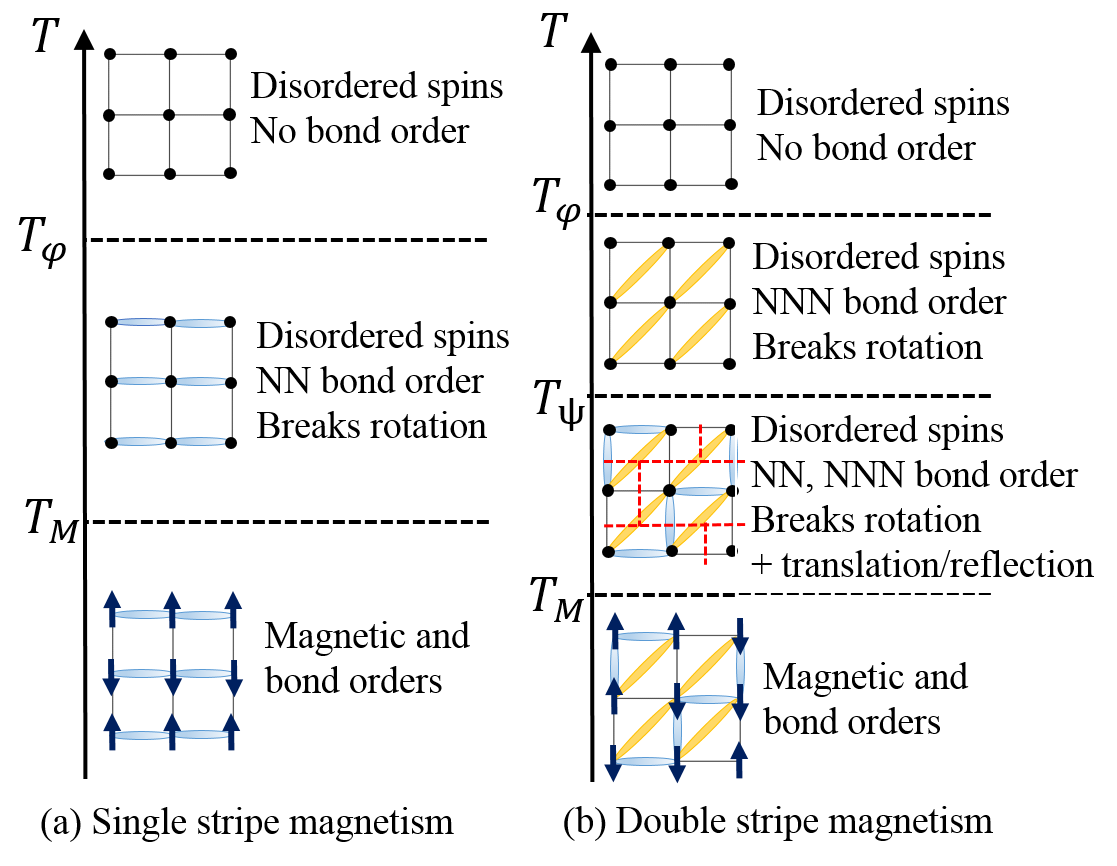}
\caption{Multi-stage melting of the magnetic order in a square lattice as
occurs in (a) single-stripe (SS) magnetism, and (b) double-stripe (DS)
magnetism. The nearest (next-nearest) neighbor ferromagnetic bonds are
indicated with blue (yellow) ovals. While in (a) there is one nematic bond-order
degree of freedom associated with rotational symmetry-breaking, in (b) two
bond-order degrees of freedom are associated with rotation, translation and
reflection symmetry breaking. The new two site unit cell associated with the
translation symmetry is indicated by the red dashed line. }
\label{fig:vertical_phase}
\end{figure}

Spin-driven nematicity is the phenomena whereby magnetic order that also breaks
discrete lattice rotational symmetries is melted by fluctuations in stages,
giving rise to a partially-melted order that preserves the spin-rotation (and
the time-reversal) symmetry but breaks some lattice rotation symmetries. In
analogy to the nematic phase of liquid crystals, which are partially-melted
smectic phases, this type of order has been dubbed electronic nematic order
\cite{Kivelson98}. This idea, initially conceived theoretically within the
framework of the 2D Heisenberg model \cite{Chandra1990_PhysRevLett_Ising}, was
propelled into the spotlight in 2008 as several groups independently proposed
it as an explanation of the split orthorhombic-magnetic transition in the newly
discovered Fe-based superconductors (FeBS) \cite{Fang2008_PhysRevB_Theory,
  Xu2008_PhysRevB_Ising}.
In these systems, the magnetic phase displays a single-stripe configuration,
characterized by spin-order with ordering vector $\mathbf{Q}=(0,\pi )$ or
$(\pi ,0)$ and a bond-order associated with the correlations of
nearest-neighbor parallel spins (see Fig.~\ref{fig:vertical_phase}a).
Consequently, in the nematic phase, spin-order is lost but the rotational
symmetry breaking bond-order is preserved, resulting in an orthorhombic
paramagnetic phase that extends above the onset of magnetic order.
Experimental signatures and theoretical implications of such a spin-driven nematicity have been widely
explored in FeBS \cite{Mazin2009_NatPhys_key, Fernandes2012,Fernandes2014, Christensen16}, and similar concepts were applied to other widely investigated systems, such as charge-driven nematicity in the cuprates \cite{Chubukov14, Kivelson214} and
tetragonal symmetry-breaking in topological Kondo insulators
\cite{Galitski15}. Nematic degrees of freedom may also play an important role
in the onset of high-temperature superconductivity, as recent experimental
\cite{Yoshizawa12, Gallais13, Fisher_arxiv} and theoretical works
\cite{Lederer15} have proposed.

While the general concept of partially-melted magnetic phases is
well-established both theoretically and experimentally, most work has focused on
the single-stripe case. How and whether more complex types of magnetic order
can also partially melt and promote novel nematic-like phases remain relatively
unexplored topics \cite{Xu2009_arXiv_Field}. Interestingly, the FeBS provides
another opportunity to investigate such ideas: while it is true that most of
these materials display single-stripe (SS) magnetic order, the Fe-based
chalcogenide FeTe exhibits a more complicated ``double-stripe'' (DS) magnetic
order \cite{Li2009_PhysRevB_Firstorder, Bao2009_PhysRevLett_Tunable}. As shown
in Fig.~\ref{fig:vertical_phase}, the DS phase has not only spin-order with
ordering vector $\mathbf{Q}=(\pi/2,\pi/2)$, but also two types of bond-order
involving nearest-neighbor (NN) and next-nearest-neighbor (NNN) parallel
spins. A natural question is whether these bond-orders can be stabilized even
in the absence of long-range magnetic order, similarly to the nematic phase in
the SS case, and whether they appear separately or at the same temperature.

In this paper, we systematically explore the bond-orders that can arise above
the onset of long-range DS magnetic order and argue that it may have been
already observed as a density-wave-type transition accompanied by an
orthorhombic distortion in the Ti-based oxypnictide BaTi$_{2}$Sb$_{2}$O and
related compounds \cite{Frandsen2014_NatCommun_Intraunitcell}. This conclusion
results from a combination of \textit{ab-initio} calculations and low-energy
field-theoretical modeling.
In particular, the model is consistent with the low-temperature orthorhombic
($Pmmm$) structure of BaTi$_{2}$Sb$_{2}$O with an accompanying intra-unit-cell
charge-density wave, \cite{Frandsen2014_NatCommun_Intraunitcell} which we also
observe using density functional theory, but only when magnetic ordering is
allowed. In contrast, distortions induced via the charge-density wave obtained in
nonmagnetic calculations either do not have the requisite $Pmmm$ symmetry or
are significantly higher in energy than the magnetic solutions. This is in
striking similarity with the FeBS where structural relaxation calculations in
the magnetic single stripe pattern also reproduce the low-temperature lattice
distortion. More importantly, the ground state magnetic order is a
double-stripe (also known as bicollinear) pattern, similar to the FeTe ground
state. We map the calculated \textit{ab-initio} energies onto an effective spin
model and by extension a corresponding low-energy field-theory, which comprises
not only exchange interactions up to third neighbors, but also four-spin
coupling up to second neighbors. To investigate the onset of bond-ordered
phases within this model, we analyze the low-energy field theory beyond mean
field to account for the role of spatial fluctuations. We find that, in
general, the DS order can melt in up to three stages, as shown in
Fig.~\ref{fig:vertical_phase}: as temperature is lowered, first NNN bond order
appears, lowering the $C_{4}$ rotational symmetry of the system down to $C_{2}$
(in BaTi$_{2}$Sb$_{2}$O this lowers the symmetry from $P4/mmm$ to $Pmmm$
\cite{Frandsen2014_NatCommun_Intraunitcell}). Upon further reduction of
temperature, there is an onset of NN bond-order, breaking the translation and
reflection symmetries of the lattice. Finally, at a lower temperature,
long-range magnetic order sets in. More generally, our work unveils the
existence of two emergent bond-order degrees of freedom in systems with DS
ground states, which may have fundamental impact on their thermodynamic
properties, including superconductivity, both in BaTi$_2$Sb$_2$O and also in the iron-chalcogenides.

\section{General properties of the double-stripe phase and its nematic phases}

\label{sec:gener-prop-double}

The phenomenon of partial melting of magnetically-ordered states, which is
ultimately behind the onset of nematic phases, is caused by long-wavelength
magnetic fluctuations (either thermal or quantum). Therefore, only
approaches that go beyond mean-field can capture this effect. Here, as
explained below in more detail, this will be achieved via a large-$N$
solution of the free energy functional for the DS state. Before we introduce
it, we first discuss the different types of bond-order that appear in the DS
ordered state, contrasting them with the standard SS ordered state.

\subsection{Brief review of single-stripe magnetism and nematicity}

\label{sec:nem-rev}

Spin-driven nematicity in SS states is most straightforwardly discussed by
means of a Heisenberg spin Hamiltonian. Following
Ref.~\onlinecite{Chandra1990_PhysRevLett_Ising}, we consider the following
Hamiltonian for classical spins on the two-dimensional square lattice
\cite{Chandra1990_PhysRevLett_Ising},
\begin{equation}
  \label{eq:1} H = J_{1}\sum_{\langle ij\rangle}\mathbf{S}_{i}\cdot\mathbf{S}_{j}+J_{2}
  \sum_{\langle\langle ij\rangle\rangle}\mathbf{S}_{i}\cdot\mathbf{S}_{j}
  -K_{1}\sum_{\langle ij\rangle}\left( \mathbf{S}_{i}\cdot\mathbf{S}
    _{j}\right) ^{2},
\end{equation}
where $J_{1}$ and $J_{2}>0$ are nearest and next-nearest neighbor exchange
couplings, and $K_{1}>0$ is the nearest-neighbor biquadratic coupling. In the
context of the Fe-pnictides, which are metals with itinerant Fe electrons, such
a model should be interpreted as an effective low-energy model to describe the
interplay between SS magnetism and nematicity. Indeed, the inappropriateness of
a purely localized approach is manifested by the fact that DFT calculations
\cite{Yaresko2009_PhysRevB_Interplay, Glasbrenner2014_PhysRevB_Firstprinciples}
not only give soft moments, but also a large biquadratic exchange $K_{1}$ as
compared to $J_{2},$ consistent with the experiment
\cite{Wysocki2011_NatPhys_Consistent}. In contrast, the order-by-disorder
mechanism of Ref.~\onlinecite{Chandra1990_PhysRevLett_Ising} gives a rather
small $K_{1}/J_{2}\sim10^{-3}$ \cite{Singh03}.

Single-stripe magnetism and the related nematicity occurs for $J_{2}>J_{1}/2$,
and is most simply understood by taking $J_{2}\gg J_{1}$, where $J_2$ leads to two decoupled
antiferromagnetic Néel sublattices. $J_{1}$ cannot
couple these two sublattices, as the exchange fields between sublattices one
and two cancel. However, the biquadratic term, $K_{1}$ requires that the spins
be collinear, leading to two degenerate ground states where the spins are
ferromagnetically correlated along either $\hat{x} $ or $\hat{y}$, and
antiferromagnetically correlated along the perpendicular direction. These two
degenerate ground states can be described by the wave-vectors $(0,\pi )$ and
$(\pi ,0)$, respectively, and break both the continuous spin-rotation symmetry,
and the discrete $C_{4}$ lattice rotation symmetry (i.e. the symmetry of a
square) down to $C_{2}$ (i.e. the symmetry of a rectangle), as shown in the
bottom left of Fig.~\ref{fig:vertical_phase}. These broken symmetries can be
captured by three different order parameters: two of them are vector N\'{e}el
order parameters, $\langle \mathbf{M}_{1}\rangle$ and
$\langle \mathbf{M}_{2}\rangle $ defined on each sublattice, and a bond-order
parameter describing the rotational symmetry breaking,
\begin{align}
  \nonumber \varphi & =\frac{1}{N_{s}}\sum_{i}\langle \mathbf{S}_{i}\cdot \mathbf{S}_{i+\hat{x}}-\mathbf{S}_{i}\cdot \mathbf{S}_{i+\hat{y}}\rangle \\
                    & =\langle \mathbf{M}_{1}\cdot \mathbf{M}_{2}\rangle,
\end{align}
where $N_{s}$ is the number of sites. Effectively, the sign of $\varphi $
describes the orientation of the ferromagnetic bonds, either along $\hat{x}$
($\varphi >0$) or along $\hat{y}$ ($\varphi <0$), while the magnitude of
$\varphi$ describes the strength of both the ferro- and antiferromagnetic
bonds.

The Mermin-Wagner theorem precludes any magnetic order at any finite temperature, in a strict two-dimensional lattice, as it breaks continuous
spin-rotation symmetry. Therefore, the N\'{e}el order parameters,
$\langle \mathbf{M}_{1}\rangle =\langle \mathbf{M}_{2}\rangle =0$. However,
$\varphi $ is a scalar (Ising) order parameter and breaks only the discrete $C_{4}$
symmetry, and so it that can, and does, condense at a finite temperature. While
$\varphi $ is called a nematic order parameter because it describes how the
magnetic fluctuations break $C_{4}$ symmetry, it can more generally be thought
of as a scalar \textit{bond order parameter} that breaks a discrete lattice
symmetry, which we shall generalize onto the case of DS magnetism. Here,
although the spins themselves are slowly fluctuating, the correlation of the
fluctuations between the two sublattices provides additional free energy gain
and generates a long-range order without breaking any continuous symmetry. In
momentum space, one can imagine that there is short range order at both
$\mathbf{Q}=(0,\pi)$ and $(\pi ,0)$ above $T_{\varphi }$, while below
$T_{\varphi }$ the fluctuations increase at one $\mathbf{Q}$ vector and
decrease at the other, thus breaking the rotational symmetry
\cite{Fernandes2012}. This has indeed been observed experimentally by neutron
scattering in the iron pnictides \cite{Dai13}.

Realistic systems will have some finite inter-layer coupling $J_{\perp}$ that
allows magnetism to develop at a temperature $T_{M}$ governed by
$\ln\left( J_{\perp}/J_{2}\right)$, at which point long-range
magnetic order will develop at the $\mathbf{Q}$ vector already chosen by
$\varphi$. For sufficiently small $J_{\perp}$, these two temperature scales can
remain separate \cite{Fang2008_PhysRevB_Theory, Xu2008_PhysRevB_Ising},
although they will typically merge for sufficiently large $J_{\perp}$
\cite{Fernandes2012, Batista11}, as the three-dimensionality reduces the role
of magnetic fluctuations.

\subsection{Double-stripe magnetism and nematicity: symmetry analysis}

\label{sec:model-intro}

\begin{figure*}[th]
\includegraphics[width=0.98\textwidth]{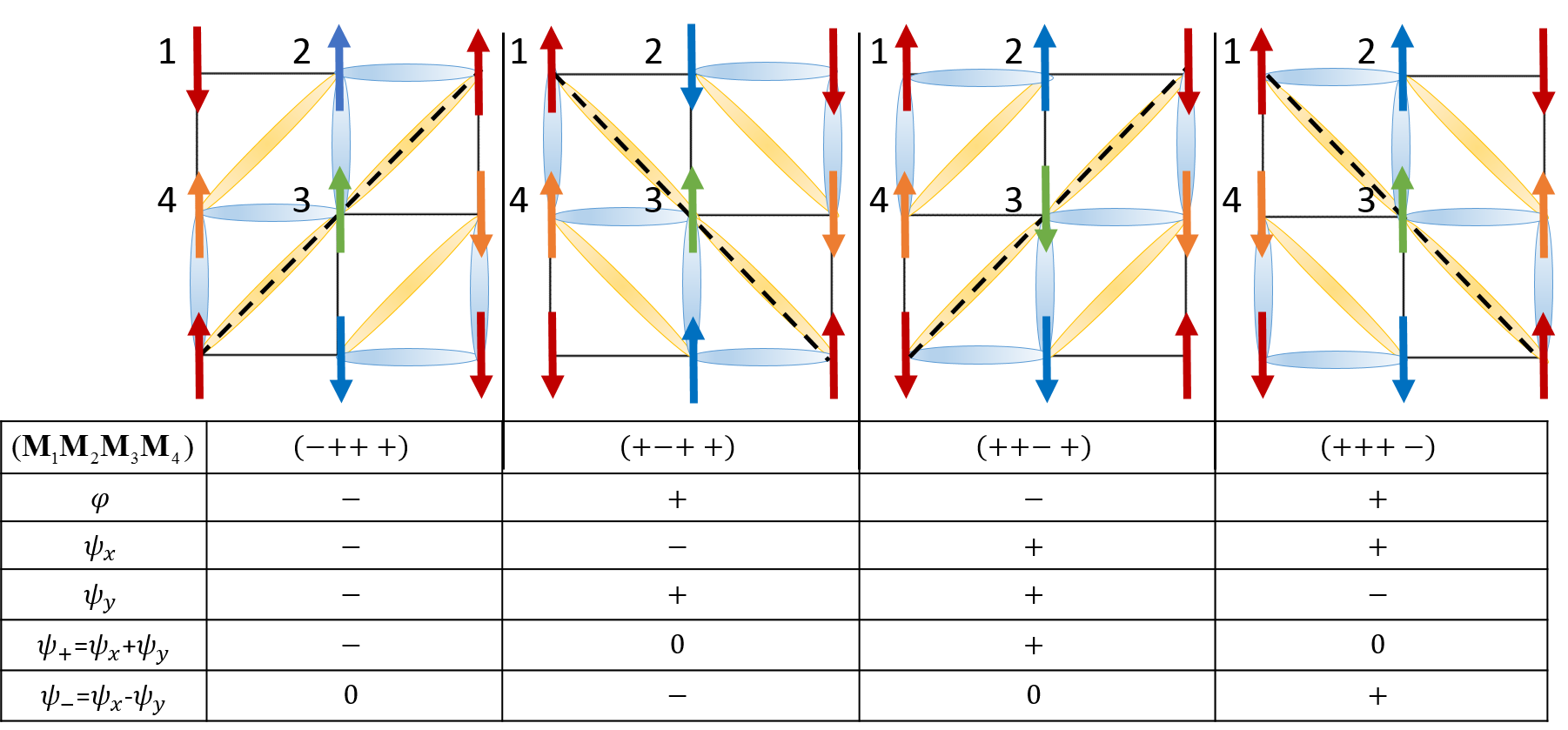}
\caption{
(Color online)The
  four degenerate ground states characterized by different configurations of
  $\mathbf{M}_{a}$'s and signs of corresponding order parameters
  $\varphi$, $\psi_{x,y}$ and $\psi_{+}$. Again the ferromagnetic bonds are indicated
  with blue/yellow ovals. The dashed black line shows the mirror plane symmetry broken by $\psi_{\pm}$. }
\label{fig:table_four}
\end{figure*}

Double-stripe magnetism consists of a plaquette of four spins -- three up, one
down, repeated with a staggered, $(\pi ,\pi )$ pattern, as shown in
Fig.~\ref{fig:vertical_phase}b, bottom panel, leading to an 4-site magnetic unit
cell (see Fig .~\ref{fig:magconfigs}(a)).   
This ordering results in double-width ferromagnetic stripes along the diagonal, alternating
antiferromagnetically, hence the name double-stripe (DS). The DS pattern can be
thought of as two copies of single-stripe orders in ``even'' and ``odd''
sublattices, rotated by $45^{\circ }$ and then coupled together by another
biquadratic coupling. In this case, the effective low-energy Hamiltonian that
displays this ground state in the classical regime is:
\begin{multline}
  \label{J1_J2_J3} H = J_{1}\sum_{\langle ij\rangle }\mathbf{S}_{i} \cdot \mathbf{S}_{j}+J_{2}\sum_{\langle \langle ij\rangle \rangle} \mathbf{S}_{i}\cdot \mathbf{S}_{j}+J_{3}\sum_{\langle \langle \langle ij\rangle \rangle \rangle } \mathbf{S}_{i}\cdot \mathbf{S}_{j} \\
  - K_{1}\sum_{\langle ij\rangle }\left( \mathbf{S}_{i} \cdot
    \mathbf{S}_{j}\right) ^{2}-K_{2}\sum_{\langle \langle ij\rangle \rangle}
  \left(
    \mathbf{S}_{i}\cdot \mathbf{S}_{j}\right) ^{2}  \\
  +R_{1}\sum_{\text{plaquette}} \left[(\mathbf{S}_{i}\cdot
  \mathbf{S}_{j})(\mathbf{S}_{k}\cdot \mathbf{S}_{l})+(\mathbf{S}_{i}\cdot
  \mathbf{S}_{l})(\mathbf{S}_{k}\cdot \mathbf{S}_{j}) \right] \\
  -R_{2}\sum_{\text{plaquette}} \left[ (\mathbf{S}_{i}\cdot
  \mathbf{S}_{k})(\mathbf{S}_{j}\cdot \mathbf{S}_{l}) \right]
\end{multline}
where $\langle \rangle $, $\langle \langle \rangle \rangle $, and
$\langle \langle \langle \rangle \rangle \rangle $ denote the first, the second
and the third nearest neighbors, respectively. The $\sum_{\text{plaquette}}$ is defined such that $ijkl$ are the indices circulating a square plaquette.  Note that the ring exchange
terms are often included with an approximation $R \equiv R_{1} = R_{2}$ (which
we also used in our DFT fits in Section \ref{sec:dens-funct-theory}), but for
itinerant systems the two coefficients can, in principle, be different.

One can understand this model by first considering the limit where $J_{3}>0$ is
the dominant interaction. If $J_{3}$ plays the pivotal role in the spin
dynamics, it is natural to partition the system into four antiferromagnetic
N\'{e}el sublattices, so that $J_{3}$ is the nearest neighbor coupling for each
of them, as shown in Fig.~\ref{fig:table_four}. Then the $J_{2}-J_{3}$ model
describes two copies of SS magnetism. The biquadratic terms, $K_{1}>0$ and
$K_{2}>0,$ force all four sublattices to be collinear. However, the DS and
plaquette (Fig.~3) orders are exactly degenerate unless the ring exchange terms
are included \cite{Glasbrenner2015_NatPhys_Effecta}. \textit{Ab-initio}
calculations (see Section \ref{sec:titan-based-oxypn}) indicate that the DS
pattern is the ground state, and also that the 4th order terms are sufficiently
strong to severely penalize noncollinear states. To simplify our analysis while
still accounting for these details, we will drop both the ring exchange terms
and only keep solutions corresponding with the symmetry of the DS ground state.  We also 
drop $J_{1},$ which generates undesirable spiral solutions that we
know are not present in our DFT calculations. 
Thus, we retain only the terms relevant to DS order, which are $J_{2,}$ $J_{3,}$
$K_{1}$, and $K_{2}$.

Besides the continuous spin-rotational symmetry, DS order also breaks a number of
discrete symmetries. One can more clearly see those discrete symmetries by
highlighting the location of the ferromagnetic bonds, as we have done in
Fig.~\ref{fig:table_four}, for the four degenerate ground states. They are: the
translational symmetry, since the unit cell is quadrupled in size; the $C_{4}$
rotational symmetry, which is broken along the diagonals of the squares
($B_{2g}$ symmetry) instead of along the sides of the square ($B_{1g} $
symmetry), as it was the case for the SS order; and the reflection symmetry
($\sigma _{d}$) across one of the diagonals ($x=\pm y$ lines). Unlike the ``broken''
translation symmetry of the single-stripe antiferromagnet, which can be
restored by a time-reversal operation (or a 180$^{\circ }$ rotation), here the
layout of the NN ferromagnetic/antiferromagnetic bonds breaks translation
symmetry and \textit{doubles} the unit cell, which is doubled again when
long-range magnetic order condenses, as shown in Fig.~\ref{fig:vertical_phase}.
In momentum space, this corresponds to 2$\mathbf{Q}$ ordering, with pairs of
$\mathbf{Q}=(\pm \pi /2,\pm \pi /2)$ that are chosen to break the rotational
symmetry appropriately.
Note that in the case of the Ti-based oxypnictides discussed in the next
section, some of those symmetries are already broken in the nonmagnetic phase
due to crystallography.

To formally describe these discrete symmetries in terms of the spins, we
consider the four N\'{e}el order parameters $\langle\mathbf{M}_{a}\rangle$ related to each of the four sublattices $a=1,2,3,4$ defined in
Fig.~\ref{fig:table_four}. We first define the two next-nearest-neighbor bond orders, which couple to
$K_{2}$:
\begin{align}
  \varphi_{\mathrm{odd}} & =\langle\mathbf{M}_{1}\cdot\mathbf{M}_{3}\rangle \\
  \varphi_{\mathrm{even}} & =\langle\mathbf{M}_{2}\cdot\mathbf{M}_{4}\rangle
\end{align}

\begin{figure}[th]
\includegraphics[width=0.48\textwidth]{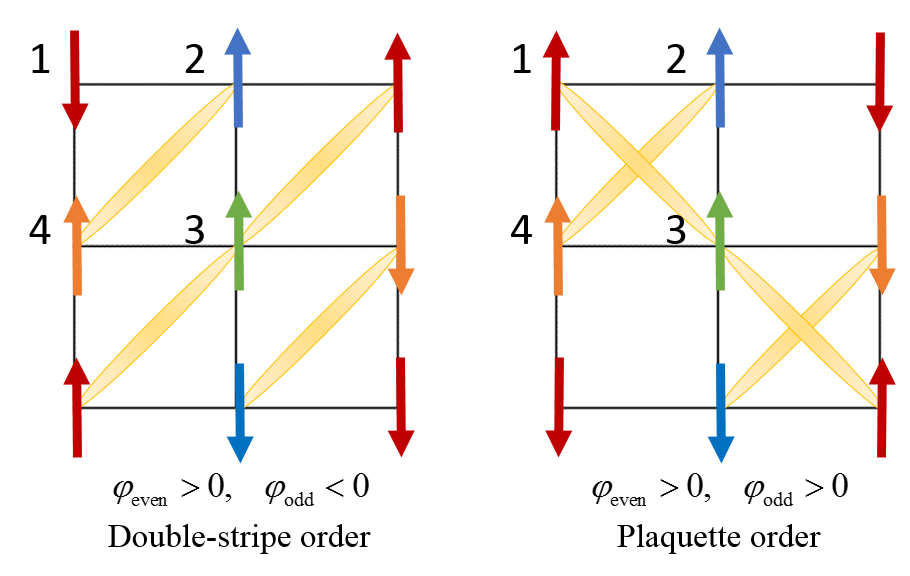}
\caption{(Color online)Signs of $\varphi_{\mathrm{even/odd}}$ in a
double-stripe order(left) and plaquette order(right). }
\label{fig:even_odd}
\end{figure}

These order parameters characterize the emergence of diagonal bond order in the
absence of long-range magnetic order, where $\varphi_{\mathrm{even/odd}}>0$
indicates which bonds within the four-spin plaquette are ferromagnetic.
Fig.~\ref{fig:even_odd} shows that when $\varphi _{\mathrm{even/odd}}$ are
opposite in sign, we can obtain the DS magnetic order, where each four-spin
plaquette has an odd number of up and down spins. In contrast, when
$\varphi _{\mathrm{even/odd}}$ have the same sign, we get the plaquette order
discussed above. Note that, while we have drawn all spins as collinear, at this
point the two sets of sublattices are decoupled and can rotate freely without
affecting the bond order. By symmetry, $\varphi _{\mathrm{even/odd}}$ must
condense at the same temperature, and indeed, it does not make sense to
condense anything but a linear combination,
$\varphi _{\mathrm{even}}\pm \varphi _{\mathrm{odd}}$, as each of
$\varphi _{\mathrm{even/odd}}$ individually does not break a well-defined
symmetry. Considering the bonds alone,
$\varphi \equiv \varphi _{\mathrm{even}}-\varphi _{\mathrm{odd}}$ breaks the
$C_{4}$ rotational symmetry, but not translation symmetry, while
$\zeta \equiv \varphi_{\mathrm{even}}+\varphi_{\mathrm{odd}}$ doubles the unit
cell, but maintains $C_{4}$ symmetry, as shown in
Fig.~\ref{fig:even_odd}. $\varphi $, of course, is consistent with DS order,
while $\zeta $ is consistent with plaquette order, and these would be
distinguished by the ring-exchange terms.

We have two reasons to believe that the DS order, and thus $\varphi$, is favored
in the real materials. First, DFT calculations for both FeTe
\cite{Glasbrenner2014_PhysRevB_Firstprinciples} and the Ti-based oxypnictides
considered in Sec.~\ref{sec:titan-based-oxypn} show that the corresponding DS
magnetic state is clearly lower in energy, which is consistent with the
experimentally observed lattice distortions in the magnetic and/or putative nematic
state. Second, $\varphi $ and $\zeta$ couple to different elastic modes and $\varphi$
may be additionally stabilized through magnetoelastic coupling
\cite{Ducatman2014_PhysRevB_Theory}. In the following, we will neglect $\zeta$
and consider only the bond orders related to DS order.

Besides next-nearest neighbor bond-order, the DS order also has
nearest-neighbor bond-orders, as shown in Fig.~\ref{fig:table_four}, which are driven by $K_{1}$. It is useful to define the generic bond-order parameter
$\psi_{ab}=\langle\mathbf{M}_{a}\cdot\mathbf{M}_{b}\rangle$ on any pair of NN
sublattices, i.e. $\psi_{12}$, $\psi_{14}$, $\psi_{23}$ and
$\psi_{34}$. However, there are only two combinations of these that are
compatible with a non-zero $\varphi$,
\begin{align}
  \nonumber \psi_{\pm} & =(\psi_{12}-\psi_{34})\pm(\psi_{14}-\psi_{23}) \\
                       & =\psi_{x}\pm\psi_{y}.
\end{align}

Each of these represents a pattern of alternating ferromagnetic and
antiferromagnetic bonds along the $x$- and $y$-axes, resulting in a
$(\pi ,\pi)$ ordering pattern that doubles the unit cell. $\psi_{x/y}$ can be
thought of as dimerization along the $x$-axis or $y$-axis respectively.
Indeed, $\psi_{\pm}$ couple to a staggered strain associated with that
dimerization of the lattice \cite{Paul2011}. This symmetry breaking is also
consistent with the intra-unit cell charge density wave observed in
BaTi$_{2}$Sb$_{2}$O \cite{Frandsen2014_NatCommun_Intraunitcell,
  Song2016_PhysRevB_Electronic}, which we will discuss further in Section
\ref{sec:titan-based-oxypn}. In addition to translational symmetry, this bond
order breaks the diagonal reflection symmetry, $\sigma_{d}$, across the line
$x=\pm y$, for $\psi_{\pm}$ respectively. Finally, it breaks the same $C_{4}$
rotation symmetry as $\varphi$. In particular, because $\psi_{\pm}$ has
ordering vector $\mathbf{Q}=\left( \pi,\pi\right) $, while $\varphi$ is a
$\mathbf{Q}=0$ order, they can only couple via a linear-quadratic combination,
i.e. $\varphi(\psi_{+}^{2}-\psi_{-}^{2})=\varphi\psi_{x}\psi_{y}$. Therefore,
as soon as $\psi_{\pm}$ develops, $\varphi$ must also turn on, but the converse
is not true.

Thus, besides the standard degeneracy related to spin-rotations, the DS ground
state has an additional fourfold degeneracy related to the scalar order
parameters $\varphi $ and $\psi _{\pm }$. These order parameters are not
independent, as discussed above and shown in Fig.~\ref{fig:table_four}: $\varphi <0$ is only compatible with $\psi_{+} \neq 0$, whereas $\varphi > 0$ is only compatible with $\psi_- \neq 0$.
Therefore, the symmetry analysis of the DS state shows that when the magnetic
ground state is the DS state, where $\varphi \neq 0$ and $\zeta =0$, we can
have two partially-melted magnetic phases: one in which only $\varphi \neq 0$,
which breaks rotational symmetry only, and another one in which both
$\varphi \neq 0$ and $\psi _{\pm }\neq 0$, which breaks rotational symmetry,
diagonal reflection symmetry, and translational symmetry. In the next section,
we use a field-theory approach to discuss the order and character of these
different transitions.

\subsection{Double-stripe magnetism and nematicity: quantitative analysis}

\label{sec:model_model}

The bond order parameters $\varphi $ and $\psi _{\pm }$ discussed above can
describe partially-melted DS phases, as long as they remain finite even in the
absence of spin order, $\left\langle \mathbf{M}_{a}\right\rangle =0$. To characterize
these phases, one needs to include magnetic fluctuations and therefore go
beyond mean-field approaches. Within the specific spin Hamiltonian
(\ref{J1_J2_J3}), this can be achieved numerically by Monte Carlo simulations
\cite{Singh03,Applegate12} or analytically by $1/S$ expansions
\cite{Chandra1990_PhysRevLett_Ising, Ducatman2012}. Here, we employ a different
approach, similarly to Ref.~\onlinecite{Fernandes2012}, that relies on a low-energy
Ginzburg-Landau free energy expansion of Eq. (\ref{J1_J2_J3}) in terms of the
four real-space Néel order parameters $\mathbf{M}_{a}$ ($a=1,2,3,4$). As
discussed above, this picture is valid in the limit where the third-neighbor
magnetic coupling $J_{3}$ is by far the largest, and has been previously
discussed for the double-stripe state \cite{Xu2009_arXiv_Field,
  Ducatman2014_PhysRevB_Theory}. The most general form of the free energy
expansion, with biquadratic exchanges taken into account, is:
\begin{multline}
  \label{S} F\left[ \mathbf{M}_{i}\right] =
  \sum_{a,b=1}^{4}\int_{\mathbf{q}}\mathbf{M}_{a,\mathbf{q}}\chi _{ab}^{-1}\left( \mathbf{q}\right) \mathbf{M}_{b,-\mathbf{q}} \\
  -\sum_{a,b,c,d=1}^{4}\int_{\mathbf{r}}\lambda _{ab,cd}\left(
    \mathbf{M}_{a}\cdot \mathbf{M}_{b}\right) \left( \mathbf{M}_{c}\cdot
    \mathbf{M}_{d}\right),
\end{multline}
The Hamiltonian (\ref{J1_J2_J3}) generates numerous $\lambda $ terms, plus,
if we allow for soft moments, as in a more itinerant model, terms with $a=b$ and/or $c=d$ are also
allowed. However, most of these are irrelevant for the $\varphi $ and $\psi $
order parameters, so we will keep only the two combinations related to the
DS order, and neglect the others. For the same reason, we will also retain
one high-symmetry term accounting for softness of the magnetic moment. Then
\begin{multline}
  \label{S_Mi} F\left[ \mathbf{M}_{i}\right] =
  \sum_{a,b=1}^{4}\int_{\mathbf{q}}\mathbf{M}_{a,\mathbf{q}}\chi _{ab}^{-1}\left( \mathbf{q}\right) \mathbf{M}_{b,-\mathbf{q}}+\frac{u}{2}\left( \sum_{a=1}^{4}\mathbf{M}_{a}^{2}\right) ^{2} \\
  -\frac{g_{1}}{2}\left( \mathbf{M}_{1}\cdot \mathbf{M}_{3}-\mathbf{M}_{2}\cdot \mathbf{M}_{4}\right) ^{2} \\
  -\frac{g_{3}}{2} \left[ \left( \mathbf{M}_{1}\cdot
      \mathbf{M}_{2}-\mathbf{M}_{3}\cdot
      \mathbf{M}_{4}\right) ^{2} \right. \\
  \left. + \left( \mathbf{M}_{1} \cdot \mathbf{M}_{4}-\mathbf{M}_{2}\cdot
      \mathbf{M}_{3}\right) ^{2}\right].
\end{multline}

The physical meaning of each term can be understood from the Hamiltonian
(\ref{J1_J2_J3}). The exchange couplings $J_{2}$ and $J_{3}$ describe the cost
of spatial fluctuations of the order parameters, and appear in the non-uniform
susceptibility $\chi _{ab}^{-1}\left( \mathbf{q}\right) $. As discussed in 
Appendix \ref{appendix_A}, in our derivation we expand $\chi _{ab}^{-1}(\mathbf{q})$ around the
ordering vector $\mathbf{Q=}\{\pi /2,\pi /2\},$ where
$\chi _{ab}^{-1}(\mathbf{Q})=r_{0}\delta _{ab},$ and $r_{0}\propto T-T_{0}$,
with $T_{0}$ denoting the mean-field magnetic transition temperature. The quadratic term in
$\mathbf{(q-Q)}$ terms are then uniquely defined by $J_{2}$ and $J_{3}$. The
$u$ term captures the cost of non-symmetry breaking longitudinal
fluctuations. Together with the first term, it defines the amplitude of the local
moments in the fully disordered case, as well as the softness of these moments.
The four spin terms between next-nearest neighbors ($K_{2}$, $R_{2}$) lead to
the $g_{1}$ term, which captures $\varphi$, while those between
nearest-neighbors ($K_{1}$, $R_{1}$) lead to the $g_{3}$ terms, which in turn
captures $\psi _{\pm }$ order.

In the mean-field approximation, the system develops DS order at $T_{0}$,
simultaneous with $\varphi $ and $\psi _{\pm }$ bond orders in a
second-order phase transition. To go beyond mean-field, we include the
effect of the long wave-length fluctuations, working in two-dimensions,
where magnetic order does not occur at any finite temperature due to the
Mermin-Wagner theorem. Here, the fluctuations suppress the magnetic
order to $T=0$. We then decouple the four quartic terms of Eq. (\ref{S_Mi}) using
Hubbard-Stratonovich transformations, which introduces four new scalar fields, 
\begin{align}
  \label{def_order_parameters1} \varphi =& g_{1}\left( \left\langle \mathbf{M}_{1}\cdot \mathbf{M}_{3}\right\rangle -\left\langle \mathbf{M}_{2}\cdot \mathbf{M}_{4}\right\rangle \right) \\
  \label{def_order_parameters2} \psi _{x}=& g_{3}\left( \left\langle \mathbf{M}_{1}\cdot \mathbf{M}_{2}\right\rangle -\left\langle \mathbf{M}_{3}\cdot \mathbf{M}_{4}\right\rangle \right) \\
  \label{def_order_parameters3} \psi _{y}=& g_{3}\left( \left\langle \mathbf{M}_{1}\cdot \mathbf{M}_{4}\right\rangle -\left\langle \mathbf{M}_{2}\cdot \mathbf{M}_{3}\right\rangle \right) \\
  \label{def_order_parameters4} \eta =& u\sum_{i=1}^{4}\langle \mathbf{M}_{i}^{2}\rangle,
\end{align}

\begin{figure}[th]
\includegraphics[width=0.48\textwidth]{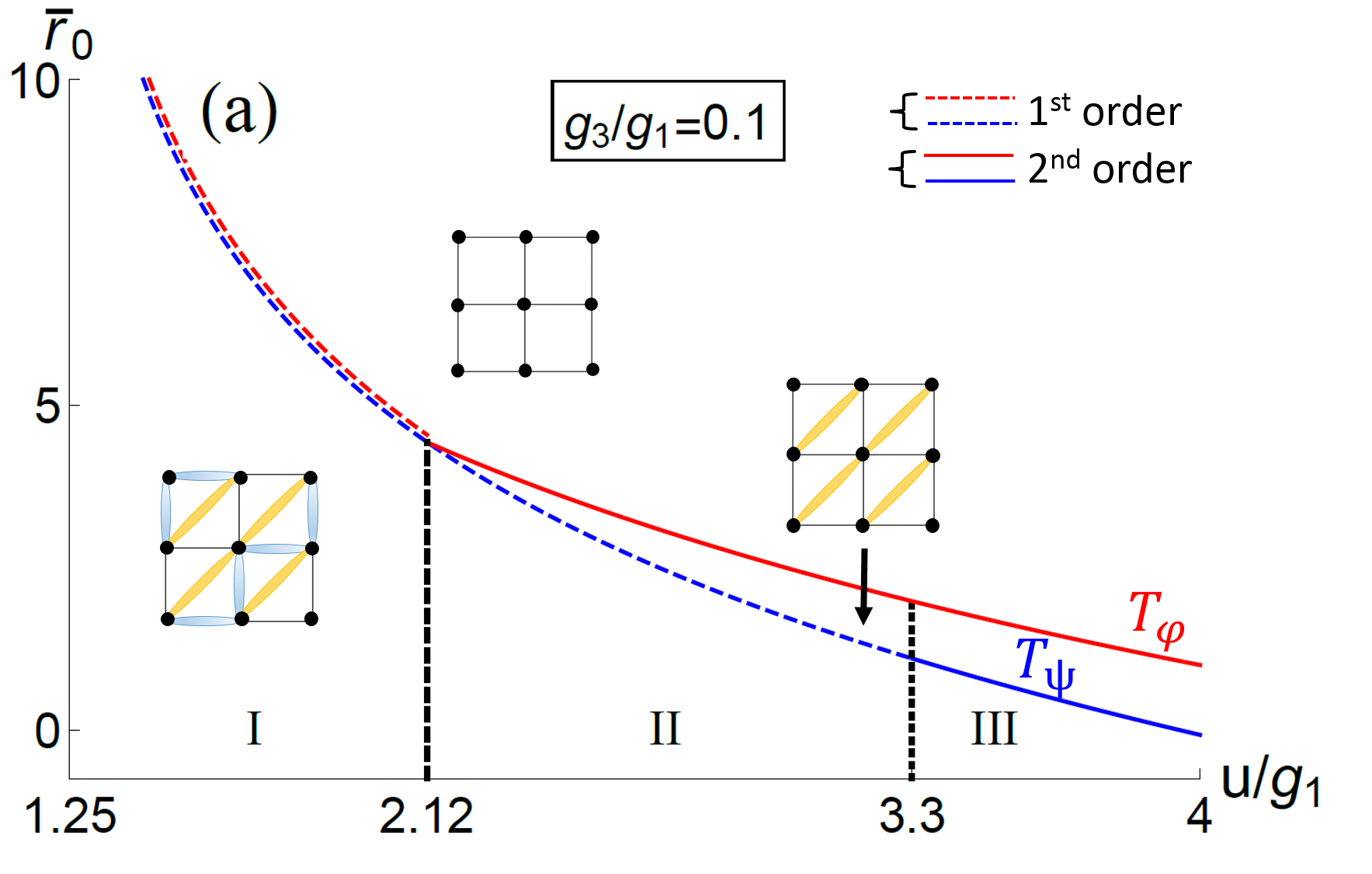}
\par
\medskip{}
\par
\includegraphics[width=0.48\textwidth]{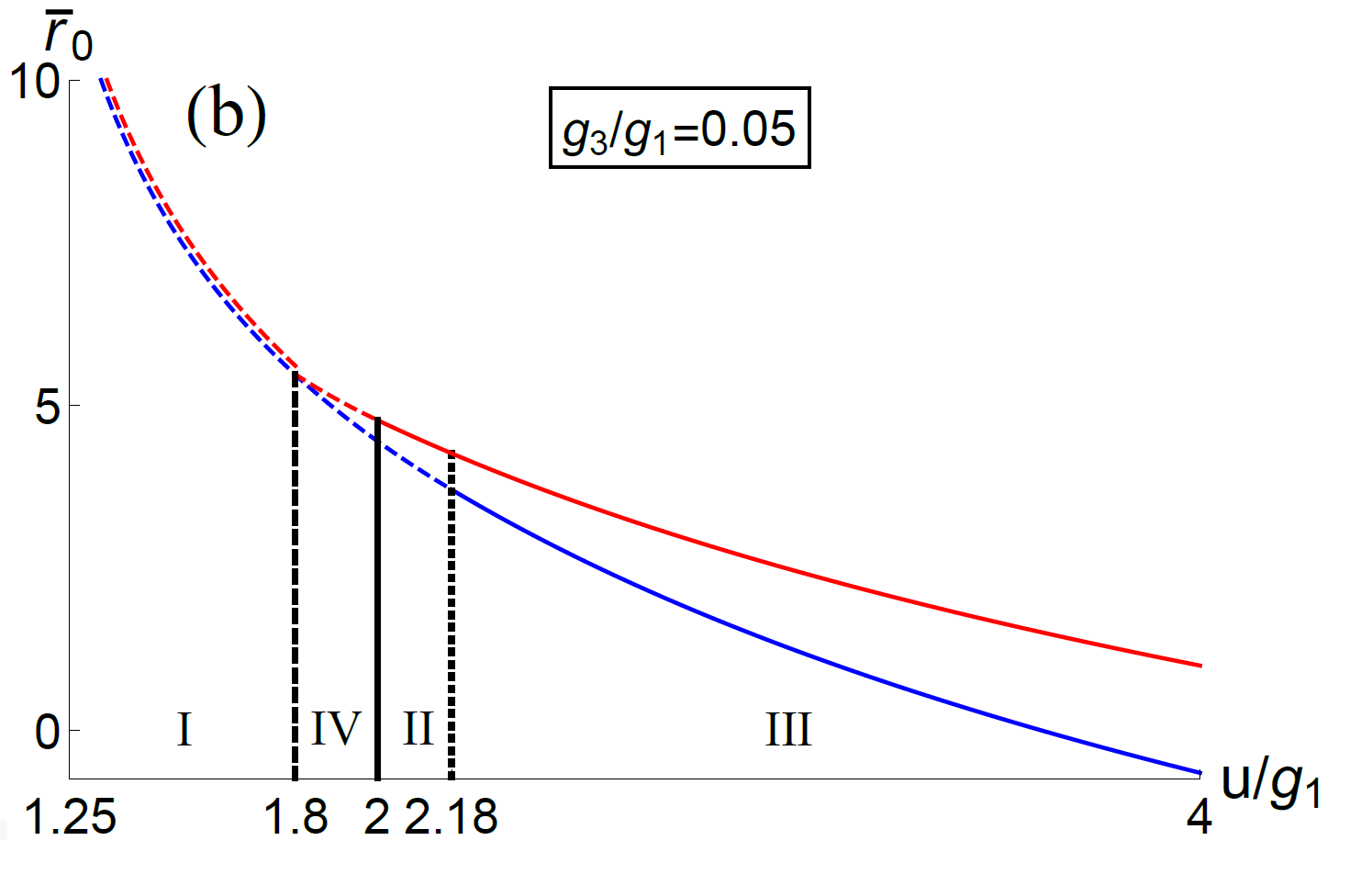}
\caption{Two examples of how $\varphi$ and $\psi_+$ orders
  develop, with the proxy for the transition temperature, $\bar{r}_{0}$ plotted
  versus $u/g_{1}$ for two values of the relative strength of the biquadratic
  terms, $g_{3}/g_{1}$. The upper, red line indicates the development of
  rotational symmetry breaking ($\varphi $), while the lower, blue line
  indicates the dimerization ($\psi_+$), which breaks the diagonal
  mirror reflection symmetry. Solid lines indicate second order transitions,
  while dashed lines indicate first order transitions, with the double-dashed
  line indicating simultaneous first order transitions. The regions of
  different classes of behavior are indicated in
  Fig~\ref{fig:phase_trans}.}
\label{figtrans}
\end{figure}

The scalar fields $\varphi $ and $\psi _{x/y}$ are equivalent to the bond order
parameters introduced in the previous subsection, and therefore break the
rotational symmetry ($\varphi $) and translational/reflectional symmetries
($\psi _{x/y}$, or $\psi _{\pm }$), and are not subject to the Mermin-Wagner theorem. On the
other hand, $\eta $ is the mean-value of the Gaussian magnetic fluctuations,
and simply renormalizes the magnetic transition temperature from its mean-field
value $T_{0}$ to the value $T_{M}$ defined via $r=r_{0}+\eta \propto
T-T_{M}$. Thus, $\eta $ is not an order parameter, as it is non-zero at any
temperature.

To proceed, we consider the two-dimensional case, where magnetic order does not
occur at any finite temperature, i.e. $\eta >-r_{0}$. In
particular, we consider the large-$N$ solution of the free energy in
Eq.~\eqref{S_Mi}, which is obtained by extending the number of components of
the $\mathbf{M}_{a}$ fields from $3$ to $N$ and taking the limit
$N\rightarrow \infty $. This yields a system of coupled self-consistent
equations for $\varphi $, $\psi _{x}$, $\psi _{y}$, and $\eta $ (see detailed
calculation in Appendix \ref{appendix_A}). An important result of this
calculation is that the first three scalar order parameters are not
independent, but coupled in the free energy expansion according to the
trilinear term, $\varphi \psi _{x}\psi _{y}.$ Furthermore, the combinations
$\psi _{\pm }=\psi _{x}\pm \psi _{y} $ decouple from the self-consistent
equations, indicating that $\psi _{x}$ and $\psi _{y}$ order
simultaneously. Consequently, non-zero $\psi _{x/y}$ necessarily gives rise to
a non-zero $\varphi $, as discussed in the previous subsection, whereas the
converse is not true.

Therefore, we define two different bond-order transition temperatures:
$T_{\varphi}$, which signals the onset of NNN bond-order $\varphi\neq0$ (with
$\psi_{\pm}=0$ and $M_{a}=0$), and $T_{\psi},$ which signals the onset of NN
bond-order $\psi_{\pm}\neq0$ (with $\varphi\neq0$ and $M_{a}=0$). Note that
whether $\psi_{+}$ or $\psi_{-}$ become non-zero depend on the sign of
$\varphi$: while $\varphi>0$ gives $\psi_{-}\neq0$, $\varphi<0$ gives
$\psi_{+}\neq0$ (see also Fig.~\ref{fig:table_four}).

In Fig.~\ref{figtrans}(a) and (b), we show two different classes of phase
diagrams.  The critical ${\bar{r}}_{0}\equiv r_{0}+8u\ln\Lambda$ (as defined
in Appendix \ref{appendix_A}), acts a proxy for temperature, and is plotted versus
$u/g_{1}$ for two representative relative strengths of the biquadratic couplings,
$g_{3}/g_{1}$. The NNN bond-order always onsets at the highest temperature,
either alone ($T_{\varphi}>T_{\psi}$), in which case the transition can be
either first or second order depending on $u/g_{1}$; or simultaneously with
$\psi_{\pm}$ ($T_{\varphi}=T_{\psi}$), in which case the double transition must
be first order. In the case $T_{\varphi}>T_{\psi}$, note that $T_{\psi}$ may be
first or second order, depending on the parameter regime. 

We can also understand these orders in momentum space, where
the magnetic fluctuation spectrum at high temperatures is isotropic, with broad peaks at all four $\mathbf{Q}=\left( \pm\pi/2,\pm\pi/2\right) $ vectors.  As the
system cools down below $T_{\varphi}$, two combinations of the
$\mathbf{Q}=\left( \pm\pi/2,\pm\pi/2\right) $ vectors develop stronger
fluctuation amplitudes than the other two combinations, breaking the rotational symmetry. Upon further cooling to below
$T_{\psi} $, the two sets of fluctuations become phase correlated.

\begin{figure}[th]
\includegraphics[width=0.48\textwidth]{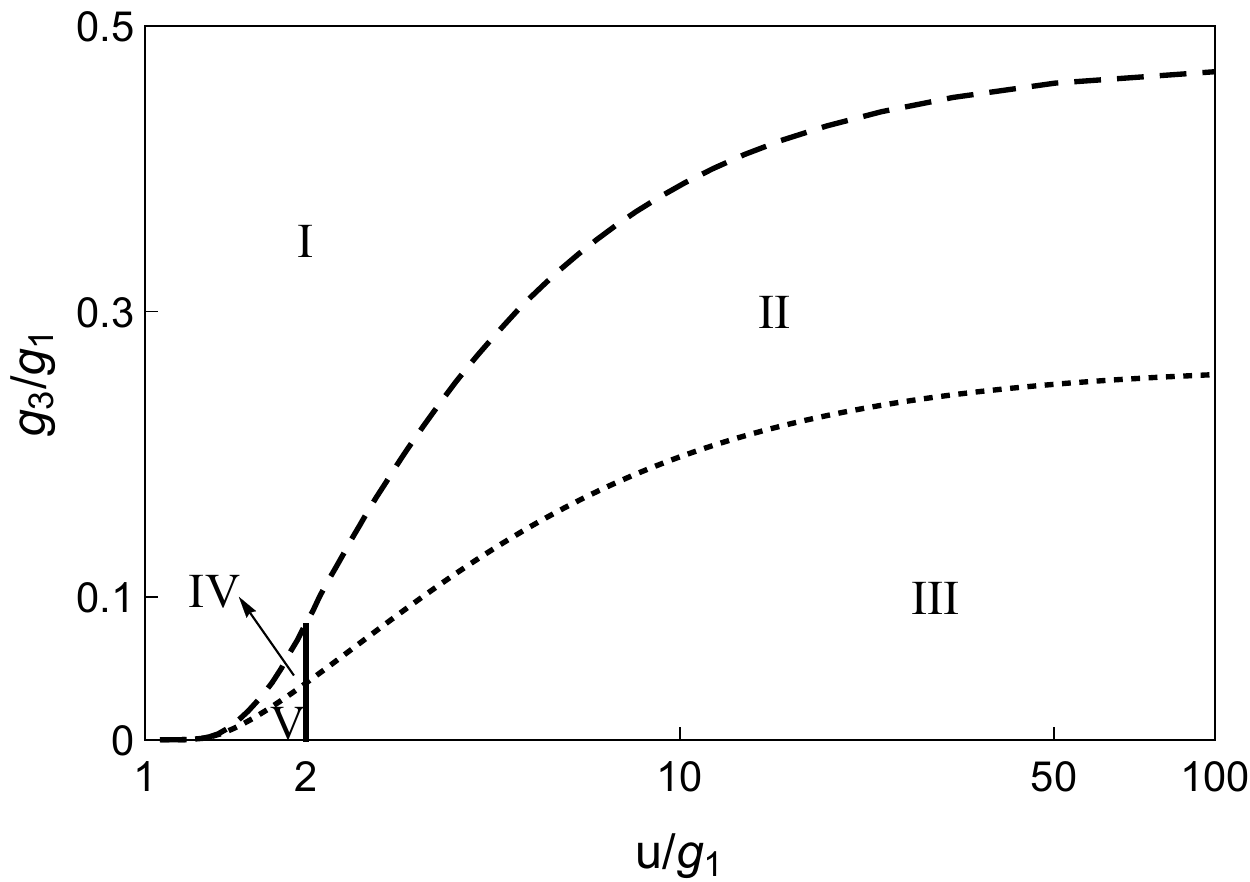}
\caption{(Color online) Classes of phase transition behavior as the
  relative strength of the biquadratic terms, $g_{3}/g_{1}$ and $u/g_{1}$ are
  varied.  \textbf{I}: Simultaneous first order transitions of $\varphi$ and
  $\psi_+$; \textbf{II}: Second order transition to $\varphi$ followed by a
  first order transition to $\psi_+$; \textbf{III}: Distinct second order phase
  transitions of $\varphi$ and $\psi_+$; \textbf{IV}: Distinct first order
  transitions of $\varphi$ and $\psi_+$; \textbf{V}: First order transition to
  $\varphi$ followed by a second order transition to $\psi_+$. }
\label{fig:phase_trans}
\end{figure}

As we have two control parameters, $u/g_{1}$ and $g_{3}/g_{1}$, we can explore
a two-dimensional phase space, as indicated in
Fig.~\ref{fig:phase_trans}. There are five different regimes of
behavior\textbf{. I}: $\varphi $ and $\psi_+ $ turn on simultaneously at a first
order transition.  \textbf{II}: $\varphi $ turns on continuously, with a second
order transition, followed by a first order transition of $\psi_+
$. \textbf{III}: two distinct second order phase transitions of $\varphi $ and
$\psi_+ $.  \textbf{IV}: two distinct first order transitions of $\varphi $ and
$\psi_+ $.  \textbf{V}: a first order transition to $\varphi $ followed by a
second order transition to $\psi_+ $. Note that these results are strongly
dependent on the two-dimensionality: any finite inter-layer coupling will genreate a finite temperature magnetic phase transition. For relatively weak couplings, the phase diagrams can be quite complicated\cite{Zhang2016}, although as the couplings approach the three-dimensional limit, all three transitions will become first order and simulataneous, and there are no pre-emptive nematic transitions,
as in the single-stripe case \cite{Fernandes2012, Fernandes2014, Chubukov2015,
  Xu2008_PhysRevB_Ising, Fang2008_PhysRevB_Theory, Abrahams2011,
  Mazin2009_NatPhys_key, Kamiya2011, Capati2011, Brydon2011, Liang2013,
  Yamase2015}. 

\section{Titanium-based oxypnictides}

\label{sec:titan-based-oxypn}

In the previous section we outlined the general theory of two-stage
spin-driven nematicity, which made no assumptions about the
chemical composition of the system. We now consider a real-world example
using \textit{ab-initio} DFT calculations that show that our model may be
realized in the Ti-based oxypnictide BaTi$_{2}$Sb$_{2}$O. We begin by
reviewing what is known experimentally about this family of materials
followed by a brief discussion of previous DFT results. We detail our
computational methods and present our calculations, which we discuss in the
context of the model. The model and DFT results provide a consistent
framework for interpreting what is known from experiment and indicates that
magnetic fluctuations drive phenomena such as the nematic phase and the
recently observed charge density wave.

\subsection{Experimental status}

\label{sec:experimental-status}

The family of Ti-based oxypnictides contains two groups of compounds,
BaTi$_{2}$Pn$_{2}$O (Pn = As, Sb, Bi) and Na$_{2}$Ti$_{2}$Pn$_{2}$O (Pn = As,
Sb). These materials share common features, such as having layered tetragonal
crystal structures similar to the Fe-based superconductors and with most
compounds also exhibiting a density wave transition (the transition is
suppressed in BaTi$_{2}$Bi$_{2}$O \cite{Yajima2012_JPhysSocJpn_Synthesis,
  Yajima2013_JPhysSocJpn_Two}). The density wave transition occurs at
$T_{DW}=50$ K for BaTi$_{2}$Sb$_{2}$O
\cite{Yajima2012_JPhysSocJpn_Superconductivity,
  Doan2012_JAmChemSoc_Ba1xNaxTi2Sb2O}, $T_{DW}=200$ K for BaTi$_{2}$As$_{2}$O
\cite{Wang2010_JPhysCondensMatter_Structure}, and $T_{DW}=330$ K and $120$ K
for the respective Na$_{2}$Ti$_{2}$Pn$_{2}$O (Pn = As, Sb) materials
\cite{Adam1990_ZAnorgAllgChem_Darstellung, Ozawa2000_JSolidStateChem_Powder,
  Ozawa2001_ChemMater_Possible}. A subset of these compounds are
superconductors, with BaTi$_{2}$Sb$_{2}$O being the prototypical example
\cite{Yajima2012_JPhysSocJpn_Superconductivity,
  Doan2012_JAmChemSoc_Ba1xNaxTi2Sb2O} with a critical superconducting
temperature of $T_{c}=1.2$ K
\cite{Yajima2012_JPhysSocJpn_Superconductivity}. Suppressing the density wave
by substituting K for Ba increases $T_{c}$ up to $T_{c}=6.1$ K
\cite{Pachmayr2014_SolStSci_Superconductivity}, meaning that, as in the Fe-based
superconductors, there is a correlation between superconductivity and the
suppression of the density wave transition.  However the critical
superconducting temperatures are much smaller, so there is interest in
understanding the differences between the Ti-based and Fe-based pnictides.

There is an active debate regarding the microscopic details and origin of the
density wave (DW) transition in the Ti-based oxypnictides that hinges on two
primary questions: 1) Is it a charge-density wave or a spin-density wave, and
2) what is the wave-vector of the DW? A set of NMR measurements, while not
being able to resolve whether or not the DW has a charge or magnetic origin
\cite{Kitagawa2013_PhysRevB_swave}, placed symmetry constraints on the DW,
finding that it broke the four-fold rotational symmetry at the Sb sites without
enlarging the unit cell, making an incommensurate DW unlikely. Neutron powder
diffraction measurements \cite{Frandsen2014_NatCommun_Intraunitcell} tightened
these constraints by detecting a lattice distortion that accompanies the DW,
changing the space group from $P4/mmm$ to $Pmmm$ due to a breaking of the
four-fold rotational symmetry, but follow-up electron diffraction measurements
did not detect a change in the number of Ti atoms per unit cell. The authors
Ref.~\onlinecite{Frandsen2014_NatCommun_Intraunitcell} identified this as a
nematic phase similar to what is observed in the Fe-based superconductors and
proposed an ``intra-unit-cell'' charge-density wave to explain their
results. This contrasts with Ref.~\onlinecite{Song2016_PhysRevB_Electronic},
where the authors claim to have detected a CDW with wave-vector
$\mathbf{Q} = (\pi, \pi)$ using angle-resolved photoemission spectroscopy and
scanning tunneling microscopy measurements. This would mean that the DW breaks
both rotational and translational symmetry and increases the unit cell size to
four Ti sites, which is incompatible with the $Pmmm$ symmetry reported in
Ref.~\onlinecite{Frandsen2014_NatCommun_Intraunitcell}. In addition, while a
long-range spin-density wave has yet to be detected in BaTi$_{2}$Sb$_{2}$O,
none of these experiments have ruled out the potential existence of magnetic
fluctuations around and below the DW transition temperature.

\subsection{Density functional theory calculations}

\label{sec:dens-funct-theory}

While experimental measurements of BaTi$_{2}$Sb$_{2}$O have yet to detect
magnetism, DFT calculations \cite{Singh2012_NewJPhys_Electronic,
  Suetin2013_JAlComp_Structural, Wang2013_JApplPhys_electronic} show a
preference for magnetism in BaTi$_{2}$Sb$_{2}$O and predict the ground state to
be the double stripe pattern. Including electronic correlations with the
DFT+U correction further stabilizes the tendency towards magnetism
\cite{Wang2013_JApplPhys_electronic}. In contrast, nonmagnetic calculations
predict a phonon instability at $\mathbf{Q} = (\pi, \pi)$ in the high
temperature structure \cite{Subedi2013_PhysRevB_Electronphonon,
  Nakano2016_arXiv_Lattice}. Similar to experiment, the DFT calculations appear
to point in multiple and exclusive directions, which complicates analysis of
the DW transition and leaves open the possibility that the superconductivity in
BaTi$_{2}$Sb$_{2}$O could be either conventional (electron-phonon coupling) or
unconventional (spin-fluctuation mediated).

Many of these conflicts observed in both theory and experiment can be equitably
resolved in our model, provided it is applicable to BaTi$_{2}$Sb$_{2}$O. To
establish this, we calculate exchange parameters using DFT calculations, which
confirms that BaTi$_{2}$Sb$_{2}$O is in the double-stripe regime described in Section
\ref{sec:model_model}. We also revisit the nonmagnetic phonon instability and
compare it with structural relaxations performed on the double stripe magnetic
state, where we observe that the double-stripe magnetic pattern calculations
yields a charge imbalance on two inequivalent Ti sites along with a
orthorhombic distortion, which is consistent with our model and also the
results of Ref.~\onlinecite{Frandsen2014_NatCommun_Intraunitcell}. We conclude
that this provides strong evidence that the DW transition corresponds to a
spin-fluctuation-driven nematic intra-unit-cell CDW that breaks four-fold rotational
symmetry.

\begin{figure}[tbp]
\centering
\includegraphics[width=0.48\textwidth]{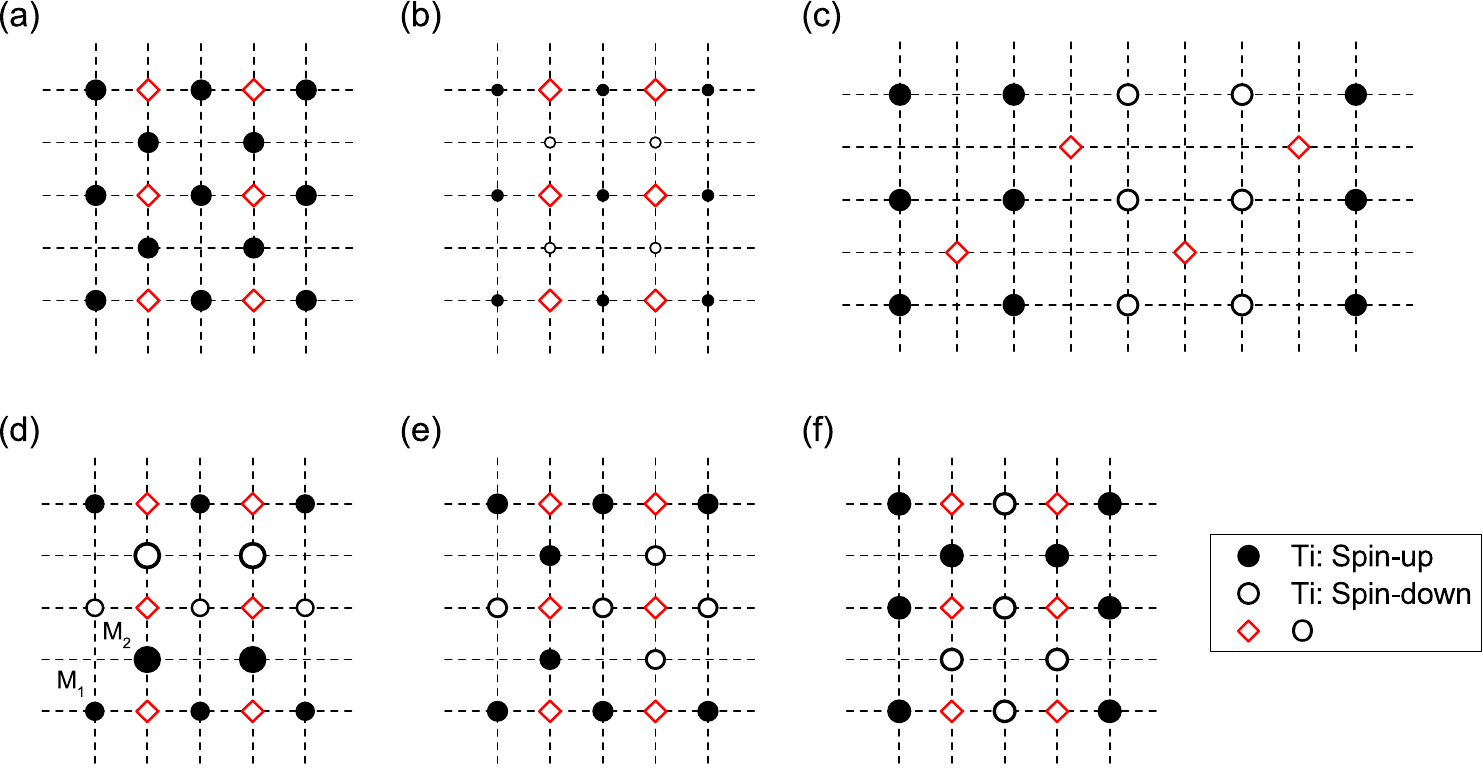}
\caption{Schematic illustrating the different magnetic patterns considered in
  our collinear calculations. The inequivalent Ti sites are labeled as $M_{1}$
  and $M_{2}$ in panel (d), and indicate NNN FM bonds bridging oxygen or
  vacancy sites, respectively. The relative sizes of the circles representing
  the Ti sites show the variation in local moment amplitudes (based on LSDA$+U$
  calculations with $U = 3.5 \text{ eV}$) across the magnetic patterns. (a)
  Ferromagnetic (FM), (b) Checkerboard (CB), (c) Parallel stripes, (d)
  Double stripes (DS), (e) Oxygen-centered plaquettes (f) Vacancy-centered
  plaquettes.}
\label{fig:colpatterns}
\end{figure}

\subsubsection{Computational methods}

\label{sec:comp-meth}

Additional details of our DFT calculations can be found in Appendix
\ref{sec:appendix_dft}. In most calculations, we used the all-electron code
\textsc{elk} \cite{elk}, with testing selected calculations against the WIEN2k
code \cite{wien2k}. For the exchange-correlation potential we used both the
local spin-density approximation (LSDA) \cite{Perdew1992_PhysRevB_Accurate} and
the generalized gradient approximation (GGA)
\cite{Perdew1996_PhysRevLett_Generalized} when computing collinear magnetic
energies. To account for correlations on Ti, we used the DFT$+U$ method in the
fully localized limit \cite{Liechtenstein1995_PhysRevB_Densityfunctional},
using two values of $U$, $2.5\text{ eV}$ and $3.5\text{ eV}$, and
$J=0.5\text{ eV}$. Due to computational expense only the LSDA$+U$ functional
with $U=3.5\text{ eV}$ was used in noncollinear calculations.

We used the experimental crystal structure in all of our calculations
\cite{Doan2012_JAmChemSoc_Ba1xNaxTi2Sb2O}. The space group symmetry is $P4/mmm$
and the lattice parameters were set to a = 4.1196 Å\ and c = 8.0951 Å. The
Wyckoff positions for the atoms, given in fractional coordinates, are: Ba [1d]
(0, 0, 0), Ti [2f] (0, 0.5, 0.5), Sb [2g] (0.5, 0.5, 0.2514), and O [1c] (0, 0,
0.5). The raw results of these calculations are presented in Appendix
\ref{sec:total-energy-calc}.

The computed DFT energies were fit to the Hamiltonian in Eq.~\ref{J1_J2_J3},
which includes the lowest-order ring exchange terms. We included this term to
capture the energy difference between the plaquette and double stripe
configurations. Note that we assumed $R_{1} = R_{2} \equiv R$ for our fits.

The crystallography of the Ti-based oxypnictides complicates the comparison
between these materials and the model described above both by breaking
symmetries and modifying exchange interactions. With regards to the exchange
interactions, the positions of the O atoms in the two-dimensional Ti$_{2}$O
plane, see for example the schematics in Fig.~\ref{fig:colpatterns}, call for
two types of NNN Ti-Ti bonds, those that are bridged by an O and those that are
not. The consequences of this are two-fold: $J_{2}$ and $K_{2}$ are split into
two unequal terms and in the DS magnetic pattern two Ti sites become
inequivalent, see Fig.~\ref{fig:colpatterns}(d).  Because of the moment
softness the local moment amplitude of one site can be smaller by a factor of
two when compared with the other (in the most extreme case the smaller moment
collapses to zero, see Appendix \ref{sec:total-energy-calc}). In principle,
this allows for two inequivalent DS patterns that differ depending on whether
the FM bonds bridging O involve either large- or small-moment Ti sites. The
moment softness also leads to different local moment amplitudes across magnetic
patterns, which is illustrated in Fig.~\ref{fig:colpatterns} by varying the
relative size of the circles, which represent Ti sites, in the plots.
While these complications are important for real Ti-based
oxypnictides and is a likely source of the crystallographic complexity of the
low-temperature phases, for simplicity we make the following assumptions when
fitting to Eq.~\eqref{J1_J2_J3}: we assume that the spins $S$ always have the
same magnitude and normalize the values of $J$'s and $K$'s to $S=1$. For
the purpose of mapping our calculations to the model described in Section
\ref{sec:model_model}, we take the average the crystallographically
inequivalent $J_{2}$'s, $K_{2}$'s and $R$'s (see Appendix
\ref{sec:comp-meth-addit} and \ref{sec:total-energy-calc}).

For the nonmagnetic and long-range antiferromagnetic configurations we also
performed structural relaxations using the projector augmented wave potentials
in the pseudopotential code \textsc{vasp} \cite{Kresse1993_PhysRevB_initio,
  Kresse1996_PhysRevB_Efficient}. One should keep in mind that, as we know from
Fe-based superconductors, the role of the long magnetic order is to break the
symmetry and create disbalance in orbital populations, which, in turn, couples
to the lattice and generate a small lattice distortion. Many calculations for
Fe pnictides and selenides show that the crystal structure in the
symmetry-broken nematic states is very well described by the corresponding
long-range ordered magnetic states, and we expect the same to be true here. In
all our relaxations we fixed the volume to the experimental value and allowed
the c/a ratio and ionic positions to relax. For the nonmagnetic instability, we
considered the vanilla GGA \cite{Perdew1996_PhysRevLett_Generalized} functional
as well as the LSDA+U and GGA+U functionals with a rotationally invariant
$U-J=3.0$ eV \cite{Dudarev1988_PhysRevB_Electronenergyloss}, and for the double
stripe relaxation we considered both LSDA+U and GGA+U with $U-J=3.0$ eV.

\subsubsection{Results and discussion}

\label{sec:results-discussion}

We checked both the LSDA and GGA functionals with Hubbard $U$ values of 3.5 eV
and 2.5 eV. GGA has more of a propensity towards magnetism, such that the
GGA+$U=3.5$ eV calculations generated too large magnetic moments, thus we did
not use this combination.  We calculated the following magnetic patterns:
ferromagnetic (FM), checkerboard (CB), double stripe (DS), oxygen- and
vacancy-centered plaquettes, parallel stripes, and single stripes
\footnote{Other patterns are possible on the two-dimensional square lattice,
  although many of them were not stable in all or some functionals. The
  staggered dimers and trimers patterns, which are competitive in bulk FeSe,
  are not stable. In addition, ferrimagnetic patterns involving 8 Ti sites with
  unequal numbers of up and down spins were also not stable}. Note that the DS
states can be converged, when $U$ is included, to two different states
differing by the local moment amplitude on the ``weak'' Ti site, which can
either stay finite or collapse to zero \cite{Yu2014_JApplPhys_siteselective}
(the relative amplitude is always smaller than the ``strong'' site). What is
important is that the symmetry breaking remains the same in both cases,
regardless of whether the ``weak'' site collapses. We also calculated the
energy of ferromagnetic planes with antiferromagnetic stacking to get an
estimate of the interplanar coupling. The energy calculations are summarized in
Appendix \ref{sec:total-energy-calc}.

The fitted values of the exchange parameters that we obtained using LSDA$+U$
with $U = 3.5 \text{ eV}$ are $J_{1} = 0.89 \text{ meV}$,
$J_{2} = -2.83 \text{ meV}$, $J_{3} = 2.79 \text{ meV}$,
$R = -0.26 \text{ meV}$, $K_{1} = -0.37 \text{ meV}$, and $K_{2} =
2.06$. The full table of fitted parameters using different functionals and
values of $U$ is available in Appendix \ref{sec:appendix_dft}. While we
note that there is noticeable variation of the absolute and even relative
values of the exchange parameters across different functional and $U$
combinations, there are important qualitative observations we can make that
hold in all cases: (1) the interaction is long-range, with
$\abs{J_{2}} > \abs{J_{1}}$ and $J_{3} > (J_{1} -|J_{2}|)/2$ (the latter
condition defines the double stripes as the mean-field ground state for
sufficiently large $K$); this is an important prerequisite for the double-stage
bond-orders described above. (2) $J_{2}$ is ferromagnetic in contrast to the
Fe-based pnictides; however this sign difference is irrelevant to the model derived in the previous section. (3) There is a sizeable biquadratic coupling,
$K_{2} > J_{3} / 2 > \abs{J_{2}} / 2 > J_{1} / 2$,
$K_{1} + 2R > J_{3} / 2 > \abs{J_{2}} / 2 > J_{1} / 2$, in which both $K_{1}$
and $K_{2}$ enforce collinearity. This ensures that spiral configurations play
a minimal role and justifies setting $J_{1}$ to 0 in Section
\ref{sec:model_model}. In addition, the distinction between $K_{1}$ and $R$ is
subtle yet important, for if the latter is omitted from the fitting, $K_{1}$
turns positive, while including $R$ yields a $K_{1}$ that is slightly negative
(in the Fe-based pnictides, including $R$ does not change the sign of
$K_{1}$). In both cases, the strong NN quartic spin interactions enforce
collinear spin patterns.

The calculated exchange parameters for LDA+U support the conclusion that
BaTi$_{2}$Sb$_{2}$O is a real-world example of the model discussed in Section
\ref{sec:model_model}, corresponding to a case where
$J_{3} \gtrsim J_{1}, J_{2}$ and $(K_{1} + 2R)$, $K_{2}$ are positive and of
the same order as the Heisenberg parameters. With this established, we now turn
to discussing how the model and DFT results describe the nature of the density
wave transition.

\begin{figure}[tbp]
\centering
\includegraphics[width=0.4\textwidth]{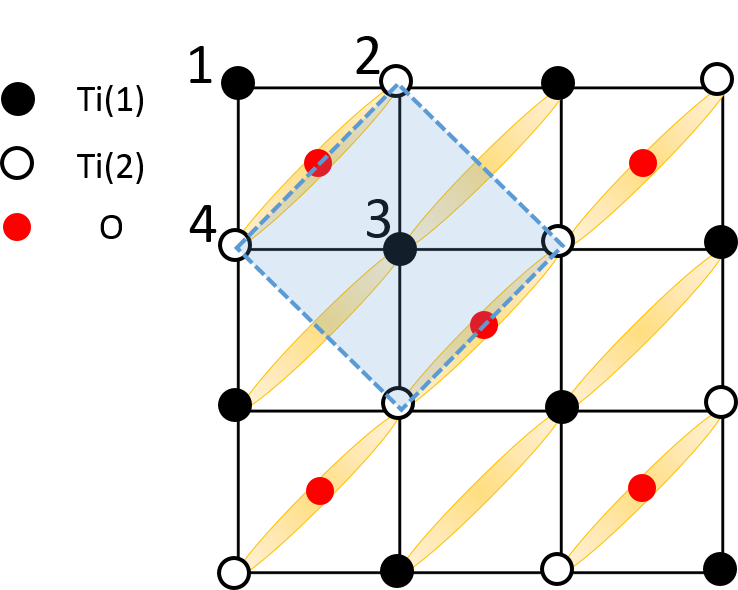}
\caption{(Color online) In BaTi$_2$Sb$_2$O, there are two Ti sites [Ti(1), hollow circles and Ti(2), 
solid black circles] per unit cell, that are differentiated by the orientation of their 
oxygen coordination (red circles).  The 2 Ti unit cell is shown by the blue, dashed lines. At high temperatures, 
these are equivalent and must carry the same charge, due to the rotation symmetry.  
However, below $T_\varphi$, the ferromagnetic bonds break rotation symmetry, and one Ti sublattice will
have ferromagnetic bonds that cross oxygen sites, while the other will not, 
allowing a charge disproportionation to develop, such that $T_\varphi = T_{CDW}$.}
\label{fig:cdw}
\end{figure}

As previously discussed, due to the crystallography of BaTi$_{2}$Sb$_{2}$O the
DFT calculation for the double stripe state has two inequivalent local Ti
moments. The DFT calculations also show a charge imbalance between the
inequivalent sites \footnote{The calculated charge in the muffin-tin spheres
  (sphere radius of 2.1 Bohr radii) at the two inequivalent Ti sites is the
  relevant quantity used here.} with there being $\sim 0.02$ more electrons at
site $M_{2}$ (site with vacancy-bridging FM NNN interactions) compared to
$M_{1}$ (site with oxygen-bridging FM NNN interactions), forming a pattern
consistent with the intra-unit-cell charge-density wave reported in
Ref.~\onlinecite{Frandsen2014_NatCommun_Intraunitcell} and in contrast to the
$\mathbf{Q} = (\pi, \pi)$ charge-density wave argued for in
Ref.~\onlinecite{Song2016_PhysRevB_Electronic}, which would lead to four inequivalent Ti sites per unit cell. 
Furthermore, our spin-driven
model is consistent with having an intra-unit-cell charge-density wave in the
absence of long-range magnetic order. The argument is as follows: the presence
of the oxygen sites breaks the translational symmetry of the hypothetical 1 Ti
tetragonal cell, such that there are 2 Ti in the primitive unit cell even for
$T > T_{\varphi}$. These Ti sites are differentiated by the direction of their oxygen
coordination, along either $\hat x \pm \hat y$, and are related by rotational symmetry.
However, when $T < T_{\varphi}$, the NNN bonds order, breaking this symmetry;
for example see Fig.~\ref{fig:cdw}. The resulting FM bonds between Ti(1)-Ti(1) and Ti(2)-Ti(2) are 
inequivalent, with one bridging an oxygen and one bridging a vacancy.  DFT calculations indicate that
this inequivalency shows up as an intra-unit-cell charge-density wave. In addition, our
model predicts that an initial unit cell with 2 inequivalent Ti sites will have
nematic order and a charge-density wave condense at the same time, in complete
agreement with experiment.  Note that the mirror symmetry associated with $\psi_\pm$ remains unbroken
until the NN bonds develop at $T_\psi$, even in the 2 Ti unit cell.

Further support for the spin-driven case comes from our structural relaxation
calculations, see Appendix \ref{sec:appendix_B} for additional details. Similar
to the Fe-based pnictides, structural relaxations of the DS pattern give rise
to an orthorhombic distortion with $Pmmm$ symmetry (consistent with
Ref.~\cite{Frandsen2014_NatCommun_Intraunitcell}). The intra-unit-cell charge
imbalance on the inequivalent Ti sites is also preserved after the
optimization. In contrast, nonmagnetic calculations in the high temperature
$P4/mmm$ structure yield a charge imbalance that resembles the double stripe
pattern. As shown in previous studies \cite{Subedi2013_PhysRevB_Electronphonon,
  Nakano2016_arXiv_Lattice}, this nonmagnetic charge density wave is unstable
and promotes one of two lattice distortions, (i) A slight rotation of the Ti
plaquettes centered around the oxygen sites as reported in
Refs.~\onlinecite{Subedi2013_PhysRevB_Electronphonon,
  Nakano2016_arXiv_Lattice}, which breaks rotation and translation symmetries
(but not the $\psi_{x/y}$ reflection symmetry) without an orthorhombic splitting of
the in-plane \textit{a} and \textit{b} lattice parameters, or (ii) an
orthorhombic distortion similar to what relaxing in the double stripe magnetic
pattern yields, which does break rotational symmetry and splits the in-plane
\textit{a} and \textit{b} lattice parameters. We stabilized both distortions in
our structural relaxations, with the former distortion being lower in energy
than the latter when using ``vanilla'' GGA or LSDA+U with $U-J=3.0\text{
  eV}$. These relaxations also remove the charge imbalance on inequivalent Ti
sites, implying that the distortions suppress the charge density wave.

It is important to emphasize that these nonmagnetic distortions are
inconsistent with experiment: the Ti plaquette rotation does not break all the
necessary symmetries, the energy of the orthorhombic distortion is higher than
the plaquette rotation and within a tenth of a meV of the undistorted
structure, and in both cases the distortion removes the charge imbalance on
inequivalent Ti sites. Both nonmagnetic distortions are also significantly
higher in energy than the magnetic double stripe configuration and its
accompanying orthorhombic distortion for the LSDA+U functional. It is only
within our spin-driven model that one obtains an orthorhombic lattice
distortion with the correct symmetry, a charge imbalance on inequivalent Ti
sites that persists after structural relaxation, and still not require that
long-range magnetic order condense. The consistency of our model in explaining
all observed phenomena also points to BaTi$_{2}$Sb$_{2}$O having a
spin-fluctuation-mediated superconducting state.

\section{Concluding remarks}

\label{sec:conclusions}

We have presented an extension to the spin-driven nematic theory that describes
fluctuations of double stripe magnetic order, which can break symmetries via a
three-stage process. The first is the formation of second nearest-neighbor
ferromagnetic bonds along one of the square diagonals, which breaks $C_{4}$
rotational symmetry, and the second is the formation of first nearest-neighbor
ferromagnetic bonds in a staggered zig-zag pattern that breaks translational
(doubling the unit cell) and reflection symmetries. Despite breaking different
symmetries, these transitions are both bond-order transitions. In principle, in a
quasi-three-dimensional system they should be followed by an antiferromagnetic
transition, but, depending on the parameters and factors that go beyond the
model, the magnetic transition may sink to an undetectable temperature. This
happens, for instance, for SS nematicity in FeSe \cite{Glasbrenner2015_NatPhys_Effecta}. 
While this seems to also be the case for BaTi$_{2}$Sb$_{2}$O, where magnetic order has not been observed experimentally,
in the DS compound FeTe the two bond-order transitions and the magnetic transition seem to be
simultaneous and first-order.
Going back to BaTi$_{2}$Sb$_{2}$O, where the magnetic transition is likely absent, the two
bond order transitions can, in general, occur at either the same or different
temperatures, depending on the relative amplitudes of the first- and
second-nearest neighbor biquadratic exchange parameters and other factors, or
the second transition may also sink to too-low temperatures. We speculate that
the former may be the case in BaTi$_{2}$Sb$_{2}$O and the resulting merged
phase transition is of very weak first order character. This would place BaTi$_{2}$Sb$_{2}$O 
in region I of the theoretical phase diagram of Fig. \ref{fig:phase_trans}. Our DFT calculations confirmed
that BaTi$_{2}$Sb$_{2}$O is within the regimes possible in this model and that
all details of existing experiments can be accounted for in the spin-driven
picture. The importance of spin fluctuations in explaining these phenomena
suggests that the superconducting state may be unconventional and driven by
spin fluctuations.

Direct confirmation of our theory should be possible with additional
measurements. We predict that the BaTi$_{2}$Sb$_{2}$O exhibits correlated
magnetic fluctuations without long-range order below the density wave
transition temperature, similar to what is observed in paramagnetic nematic
phases of specific iron pnictide superconductors, for example
BaFe$_{2}$As$_{2}$. Techniques such as muon spin rotation, which have not found
any evidence for magnetism in the titanium-based oxypnictides, are slow probes
on the time-scale of magnetic fluctuations. Fast-probe techniques such as
inelastic magnetic neutron scattering \cite{Diallo2010_PhysRevB_Paramagnetic},
photoemission spectroscopy \cite{Vilmercati2012_PhysRevB_Itinerant}, and x-ray
emission spectroscopy \cite{Gretarsson2011_PhysRevB_Revealing} measurements are
necessary to detect these fluctuations, as they have in the iron-based
superconductors. A successful detection would provide direct evidence
concerning the validity of our model. In addition, the model may also apply to
other members of the titanium oxypnictide family, such as explaining the two
phase transitions at $T = 320 \text{ K}$ (density wave)
\cite{Ozawa2001_ChemMater_Possible} and $150 \text{ K}$ (breaking of rotational
symmetry) \cite{Chen2016_arXiv_New} in Na$_{2}$Ti$_{2}$As$_{2}$O. Additional
(magnetic) DFT calculations and experiments searching for magnetic fluctuations
in Na$_{2}$Ti$_{2}$As$_{2}$O are therefore needed.

\begin{acknowledgments}
I.I.M. acknowledges Funding from the Office of Naval Research (ONR) through
the Naval Research Laboratory's Basic Research Program. J.K.G. acknowledges
the support of the NRC program at NRL. R.M.F. is supported by the U.S.
Department of Energy, Office of Science, Basic Energy Sciences, under award
number DE-SC0012336. This research was supported in part by Ames Laboratory
Royalty Funds and Iowa State University startup funds (G.Z and R.A.F.). The
Ames Laboratory is operated for the U.S. Department of Energy by Iowa State
University under Contract No. DE-AC02-07CH11358. R.A.F and R.M.F. also
acknowledge the hospitality of the Aspen Center for Physics, supported by
National Science Foundation Grant No. PHYS-1066293 where we initiated this
project.
\end{acknowledgments}

\appendix

\section{Derivation of equations of state in effective field theory}

\label{appendix_A}

In this appendix, we show briefly how to get the equations of state. After
introducing the Hubbard-Stratonovich fields in
eqns.~\eqref{def_order_parameters1}--\eqref{def_order_parameters4}, the free
energy in Eq.~\eqref{S_Mi} becomes:
\begin{multline}
  \label{eq4} F_{\mathrm{eff}}\left[
    \mathbf{M}_{i},\psi_{x/y},\varphi,\eta\right] = \\
  \sum_{i,j=1}^{4}\int_{\mathbf{q}}\mathbf{M}_{i,\mathbf{q}}\chi_{ij}
  ^{-1}\left( \mathbf{q}\right) \mathbf{M}_{j,-\mathbf{q}} -\varphi\left(
    \mathbf{M}_{1}\cdot\mathbf{M}_{3}-\mathbf{M}_{2} \cdot
    \mathbf{M}_{4}\right) \\
  -\psi_{x}\left( \mathbf{M}_{1}\cdot\mathbf{M}
    _{2}-\mathbf{M}_{3}\cdot\mathbf{M}_{4}\right) -\psi_{y}\left(
    \mathbf{M}_{1}\cdot\mathbf{M}_{4}-\mathbf{M}_{2} \cdot
    \mathbf{M}_{3}\right) \\
  +\eta\left( \sum_{i=1}^{4}\mathbf{M}_{i} ^{2}\right)
  +\frac{\varphi^{2}}{2g_{1}}+\frac{\psi_{x}^{2}}{2g_{3}}+\frac{\psi_{y}^{2}}{2g_{3}}-\frac{\eta^{2}}{2u}.
\end{multline}


Upon integrating out the $\mathbf{M}_{i}$, we obtain 
\begin{multline}
  \label{final2_Seff} F_{\mathrm{eff}}[\psi_{x,y},\varphi,\eta] = \\
  \frac{T}{2} \int_{\mathbf{q}} \log\det\mathcal{G}_{\mathbf{q}}^{-1}+\frac{\varphi^{2}}{2g_{1} }+\frac{\psi_{x}^{2}+\psi_{y}^{2}}{2g_{3}}-\frac{\eta^{2}}{2u}
\end{multline}
with $\mathcal{G}_{\mathbf{q}}^{-1}$ given by: 
\begin{equation}
\begin{bmatrix}
  (r+J_{3}\delta q^{2})\mathbb{I}-\frac{\psi_{x}}{2}\sigma_{1} & -\frac {i\psi_{y}}{2}\sigma_{2}\!-\!(J_{2}\delta q_{x}\delta q_{y}\!+\!\frac{\varphi}{2})\sigma_{3}\!\! \\ 
  \frac{i\psi_{y}}{2}\sigma_{2}\!-\!(J_{2}\delta q_{x}\delta q_{y} \!+\!\frac{\varphi}{2})\sigma_{3}\!\!\! & (r+J_{3}\delta q^{2})\mathbb{I}+\psi_{x}/2\sigma_{1}
\end{bmatrix}
\end{equation}
where $r\equiv r_{0}+\eta$.

The determinant of the inverse Green's function is: 
\begin{align}
  \nonumber \det\mathcal{G}^{-1} & = \frac{1}{16}\left( 2\tilde{J}_{2}+2\tilde{J}_{3}+2r+\varphi-\psi_{x}-\psi_{y}\right) \\
  \nonumber & \times\left( 2\tilde{J}_{2}-2\tilde{J}_{3}-2r+\varphi+\psi_{x}-\psi_{y}\right) \\
  \nonumber & \times\left( 2\tilde{J}_{2}-2\tilde{J}_{3}-2r+\varphi-\psi_{x}+\psi_{y}\right) \\
  \nonumber & \times\left( 2\tilde{J}_{2}+2\tilde{J}_{3}+2r+\varphi+\psi_{x}+\psi_{y}\right) \\
  \nonumber & = \left(\tilde{J}_{2}-\tilde{J}_{3}-\lambda\right)^{2}\left(\tilde{J}_{2}+\tilde{J}_{3}+\lambda\right)^{2} \\
  \nonumber & +2\tilde{J}_{2}\left(\tilde{J}_{2}-\tilde{J}_{3}-\lambda\right)\left(\tilde{J}_{2}+\tilde{J}_{3}+\lambda\right)\varphi \\
  \nonumber& +\frac{1}{2}\left(3\tilde{J}_{2}^{2}-(\tilde{J}_{3}+\lambda)^{2}\right)\varphi^{2}+\frac{\tilde{J}_{2}}{2}\varphi^{3}+\frac{\varphi^{4}}{16}\cr
  \nonumber & -\frac{1}{2}\left(\tilde{J}_{2}^{2}+(\tilde{J}_{3}+\lambda)^{2}\right)\left(\psi_{x}^{2}+\psi_{y}^{2}\right) \\
  \nonumber & +2\tilde{J}_{2}(\tilde{J}_{3}+\lambda)\psi_{x}\psi_{y}+\frac{1}{16}\left(\psi_{x}^{2}-\psi_{y}^{2}\right)^{2}\\
  \nonumber & +(\tilde{J}_{3}+\lambda)\varphi\psi_{x}\psi_{y} +\frac{\tilde{J}_{2}}{2}\varphi\left(\psi_{x}^{2}+\psi_{y}^{2}\right) \\
  \label{eq:inv_determinant} & -\frac{1}{8}\varphi^{2}\left( \psi_{x}^{2}+\psi_{y}^{2}\right)
\end{align}
where we have introduced $\tilde{J}_{3}=J_{3}\delta q^{2}$ and
$\tilde{J} _{2}=J_{2}\delta q_{x}\delta q_{y}$ for simplicity. In the Landau theory, we can expand the $\log\det\mathcal{G}^{-1}$ by assuming that everything involving $\varphi$, $\psi_{x}$ and $\psi_{y}$ is small in comparison to the first term. By doing so, we get a new Landau theory in terms of $\varphi$ and $\psi_{x/y}$. Once we do this expansion, we see that $\sum_{\mathbf{q}}\tilde{J}_{2}^{2n+1}$ type terms vanish. So the linear and cubic $\varphi$ terms vanish, as the $\varphi(\psi_{x}^{2}+\psi_{y}^{2})$ and $\psi_{x}\psi_{y}$ term. However, the $\varphi\psi_{x}\psi_{y}$ term is really there, as we expected. Since $\psi_{x}\psi_{y}$ acts like an external field for $\phi$, so either $\varphi$ turns on first, or $\psi_x, \psi_y$ and $\varphi$ all turn on at the same time.

The next step is to minimize the effective action with respect to $\eta$, $\varphi$,
 $\psi_{x}$ and $\psi_{y}$ by taking the derivative of $S_{\mathrm{eff}}[\psi_{x}, \psi_y, \varphi, \eta]$ over $\psi_{x}$, $\psi_{y}$, $\varphi$ and $\eta$ respectively and force it to be zero. It is convenient to rewrite the action as:

\begin{align}
&  S_{\mathrm{eff}}\left[\psi_{x},\psi_{y},\varphi,\eta\right]  =  \frac{\varphi^{2}}{2g_{1}}+\frac{\psi_{x}^{2}}{2g_{3}} +\frac{\psi_{y}^{2}}{2g_{3}}-\frac{ \eta^2}{2 u} \cr
&  \quad+\frac{T}{2}\sum_{\mathbf{q}}\log\left(J_{3}q^{2}+J_{2}q_{x}q_{y}+r+\varphi-\psi_{x}-\psi_{y}\right)\cr
 &  \quad +\frac{T}{2}\sum_{\mathbf{q}}\log\left(J_{3}q^{2}-J_{2}q_{x}q_{y}+r-\varphi-\psi_{x}+\psi_{y}\right)\cr
 &	\quad +\frac{T}{2}\sum_{\mathbf{q}}\log\left(J_{3}q^{2}-J_{2}q_{x}q_{y}+r-\varphi+\psi_{x}-\psi_{y}\right)\cr
 &  \quad +\frac{T}{2}\sum_{\mathbf{q}}\log\left(J_{3}q^{2}+J_{2}q_{x}q_{y}+r+\varphi+\psi_{x}+\psi_{y}\right) \qquad
\end{align}
where we renormalize $\left(\varphi,\psi_{x},\psi_{y}\right)\rightarrow2\left(\varphi,\psi_{x},\psi_{y}\right)$
and $g_{i}\rightarrow4g_{i}$ for convenience. The saddle point equations
$\frac{\partial S_{\mathrm{eff}}[x_{i}]}{\partial x_{i}}=0$($x_i = \eta, \varphi, \psi_x$ and $\psi_y$) become:
\begin{align}
  \eta & = \frac{Tu}{2}\sum_{\mathbf{q}}\left[ I_{1}\left( \mathbf{q}\right)
        +I_{2}\left( \mathbf{q}\right) +I_{3}\left( \mathbf{q}\right) +I_{4}\left( 
        \mathbf{q}\right) \right]  \notag \\
  \varphi & = \frac{Tg_{1}}{2}\sum_{\mathbf{q}}\left[ -I_{1}\left( \mathbf{q}\right) +I_{2}\left( \mathbf{q}\right) +I_{3}\left( \mathbf{q}\right)
            -I_{4}\left( \mathbf{q}\right) \right]  \notag \\
  \psi_{x} & = \frac{Tg_{3}}{2}\sum_{\mathbf{q}}\left[ I_{1}\left( \mathbf{q}\right) +I_{2}\left( \mathbf{q}\right) -I_{3}\left( \mathbf{q}\right)
             -I_{4}\left( \mathbf{q}\right) \right]  \notag \\
  \psi_{y} & = \frac{Tg_{3}}{2}\sum_{\mathbf{q}}\left[ I_{1}\left( \mathbf{q}\right) -I_{2}\left( \mathbf{q}\right) +I_{3}\left( \mathbf{q}\right)
             -I_{4}\left( \mathbf{q}\right) \right]  \label{aux_sp}
\end{align}
where $I_{l}(\boldsymbol{\mathbf{q}})(l=1,2,3,4)$ represents 
\begin{align}
I_{1}\left( \mathbf{q}\right) & = \frac{1}{Jq^{2}+r+\varphi-\psi_{x}
-\psi_{y}}  \notag \\
I_{2}\left( \mathbf{q}\right) & = \frac{1}{Jq^{2}+r-\varphi-\psi_{x}
+\psi_{y}}  \notag \\
I_{3}\left( \mathbf{q}\right) & = \frac{1}{Jq^{2}+r-\varphi+\psi_{x}
-\psi_{y}}  \notag \\
I_{4}\left( \mathbf{q}\right) & = \frac{1}{Jq^{2}+r+\varphi+\psi_{x}
+\psi_{y}}  \label{aux_aux_sp_2}
\end{align}
In this, we have rotated our momentum axes to define the effective kinetic
term $Jq^{2}=\sqrt{J_{3}^{2}-\frac{J_{2}^{2}}{4}}(\delta q_{x}^{2}+\delta
q_{y}^{2})$ and renormalized $\left( \varphi,\psi_{x},\psi_{y}\right)
\rightarrow2\left( \varphi,\psi_{x},\psi_{y}\right) $ and $g_{i}
\rightarrow4g_{i}$ for convenience.

In the spirit of Landau theory, we next approximate $T$ by $T_{0}$
everywhere except in $r_{0}$, which we assume to be relatively small. We can
then proceed to solve these equations in two dimensions by evaluating the
momentum integrals $I_{l}(\mathbf{q})$ directly. These integrals diverge in
the ultraviolet, so we must introduce a cutoff $\Lambda$. We can then absorb
this into $\bar{r}_{0}=r_{0}+8u\ln\Lambda$. As the equations for $\psi_{x}$
and $\psi_{y}$ differ only by signs, we can decouple their equations by
defining $\psi_\pm=\psi_{x}\pm\psi_{y}$, which have identical equations. The
trilinear term $\varphi\psi_{x}\psi_{y}$ becomes $\varphi
(\psi_+^{2}-\psi_-^{2})$. As only $\psi_+$ or $\psi_-$ will develop,
depending on the sign of $\varphi$, we consider only $\psi_+ $ ($\varphi<0$)
and obtain the three saddle point equations, 
\begin{align}
\frac{\bar{r}_{0}-r}{u} & = \ln(r+\varphi-\psi_+)+\ln(r+\varphi+\psi_+
)+2\ln(r-\varphi)  \notag \\
\frac{\varphi}{g_{1}} & = \ln(r+\varphi-\psi_+)+\ln(r+\varphi+\psi_+
)-2\ln(r-\varphi)  \notag \\
\frac{\psi_+}{2g_{3}} & = -\ln(r+\varphi-\psi_+)+\ln(r+\varphi+\psi_+)
\end{align}
where we have rescaled $\frac{T_{0}}{2 J^2}(u,g_{1},g_{3})\rightarrow
(u,g_{1},g_{3})$ and $\frac{1}{J}\left( \varphi, \psi_+, r \right)
\rightarrow \left( \varphi, \psi_+, r \right)$ and absorbed a pre-factor $1/(4\pi)$ into the temperature $T_0$.

\section{DFT calculations}
\label{sec:appendix_dft}

\begin{figure*}
\centering
\includegraphics[width=0.98\textwidth]{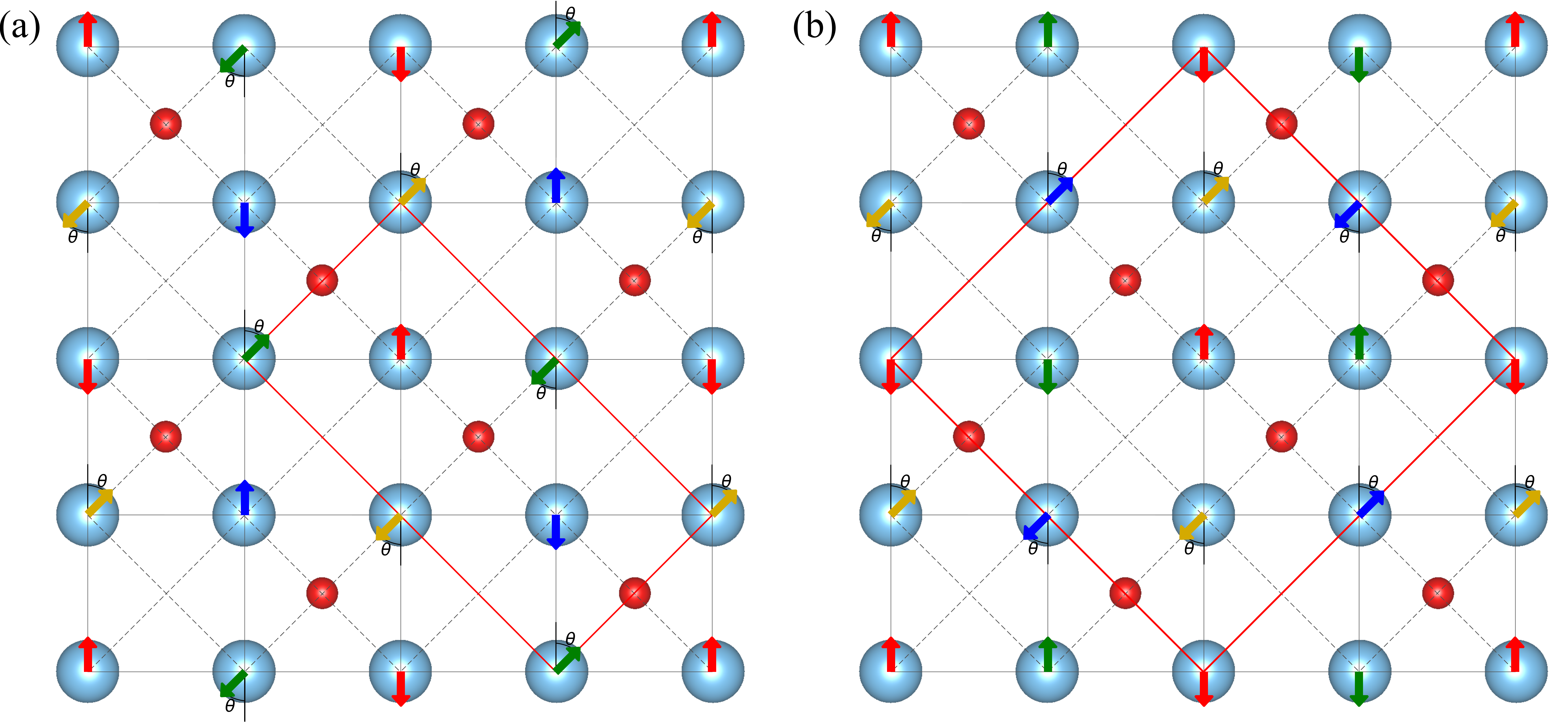}
\caption{Schematics illustrating the four interpenetrating N\'{e}el sublattices
  (shown as different color vectors) that form the double stripe magnetic
  configuration and the two different $0 \leq\theta\leq180^{\circ}$
  noncollinear rotations used in our DFT calculations to obtain the biquadratic
  terms $K_{1}$ and $K_{2}$. Red lines indicate the dimensions of the lateral
  supercell in each calculation. (a) Rotation analogous to a fluctuation of the
  $\psi_{\pm}$ order parameter (see Fig.~\ref{fig:table_four}). (b) Rotation
  analogous to a fluctuation of the $\varphi$ order parameter (see
  Fig.~\ref{fig:table_four}), which rotates between the
  $\mathbf{Q} = (\pi/2, \pi/2)$ and $(-\pi/2, \pi/2)$ configurations. }
\label{fig:magconfigs}
\end{figure*}

\begin{table*}
\centering
\begin{tabular}{cccccccccc}
\toprule & \multicolumn{6}{c}{LSDA+$U$} & \multicolumn{3}{c}{GGA+$U$} \\ 
& \multicolumn{3}{c}{U = 2.5 eV} & \multicolumn{3}{c}{U = 3.5 eV} & 
\multicolumn{3}{c}{U = 2.5 eV} \\ 
\midrule Config & E - E(NM) & $M_{1}$ & $M_{2}$ & E - E(NM) & $M_{1}$ & $M_{2}$ & E - E(NM) & $M_{1}$ & $M_{2}$ \\ 
& (meV/Ti) & \multicolumn{2}{c}{$(\mu_{B})$} & (meV/Ti) & \multicolumn{2}{c}{$(\mu_{B})$} & (meV/Ti) & \multicolumn{2}{c}{$(\mu_{B})$} \\ 
\midrule FM & 0.02993 & 0.2165 & 0.2165 & -9.668 & 0.3761 & 0.3761 & -21.761 & 0.4511 & 0.4511 \\
CB & -0.2748 & 0.01661 & 0.01661 & -13.241 & 0.1102 & 0.1102 & -28.414 & 0.1648 & 0.1648 \\
DS & -5.959 & 0.2631 & 0.3426 & -20.950 & 0.3021 & 0.5537 & -31.527 & 0.3452 & 0.5939 \\ 
DS (O-FM only) & -3.674 & 0.3220 & 0.0000 & -12.158 & 0.4783 & 0.0000 & -16.642 & 0.5383 & 0.0000 \\ 
DS (v-FM only) & -4.188 & 0.0000 & 0.4103 & -18.291 & 0.0000 & 0.6167 & -24.915 & 0.0000 & 0.6667 \\ 
Parallel & -0.9796 & 0.1752 & 0.1752 & -10.468 & 0.3891 & 0.3891 & -22.384 & 0.4363 & 0.4363 \\
Plaquette (O-centered) & -4.572 & 0.2503 & 0.2503 & -16.022 & 0.3653 & 0.3653 & -26.580 & 0.4015 & 0.4015 \\ 
Plaquette (v-centered) & -4.269 & 0.3044 & 0.3044 & -17.856 & 0.4535 & 0.4535 & -29.084 & 0.4989 & 0.4989 \\
AFM Layers & -0.6150 & 0.09566 & 0.09566 & -9.154 & 0.3890 & 0.3890 & -22.803 & 0.4866 & 0.4866 \\ 
\bottomrule
\end{tabular}
\caption{A summary of the magnetic energies for the collinear magnetic patterns
  depicted in Fig.~\ref{fig:colpatterns} using LSDA and GGA functionals and
  Hubbard $U$'s of 2.5 eV and 3.5 eV. The reported energies are referenced
  against the nonmagnetic (NM) state. For the DS patterns, the $M_{1}$ sites
  are those with NNN FM bonds bridging oxygen and $M_{2}$ are NNN FM bonds
  bridging vacancies. The ``O-FM'' and ``v-FM'' versions of DS are special
  cases when half the sites are nonmagnetic, with the magnetic sites either
  bridging across an oxygen site or a vacancy respectively. The plaquette
  patterns are labeled similarly, depending on whether they center around
  oxygen or vacancy sites.}
\label{tab:collinear_energy}
\end{table*}

In this Appendix we give additional details about how we performed the DFT
calculations and also provide the raw results of our calculations along with
additional discussion.

\subsection{Computational methods: additional details}
\label{sec:comp-meth-addit}

A method similar to that used in
Ref.~\onlinecite{Glasbrenner2014_PhysRevB_Firstprinciples} was employed for the
noncollinear calculations that were used to extract the biquadratic interaction
term. The configurations depicted in Figs.~\ref{fig:magconfigs}(a) and
\ref{fig:magconfigs}(b) indicate how we varied $\theta$ to calculate
$E(\theta)$ for the two different setups. These involve rotations of the four
N\'{e}el sublattices discussed in Section \ref{sec:model-intro}, and in our
calculations two of the sublattices are fixed and the other two are rotated to
interpolate between degenerate double stripe configurations. Applying
Eq.~\eqref{J1_J2_J3} to these configurations results in the following two
expressions that we use for fitting:
\begin{align}
  \label{eq:dft_nc_rot1}
  E_{1}(\theta) - E_{1}(0) &= 2 \left( K_{1} + 2 R \right) \sin^{2}\theta \\
  \label{eq:dft_nc_rot2}
  E_{2}(\theta) - E_{2}(0) &= \left( K_{1} + 2 K_{2} \right) \sin^{2}\theta.
\end{align}
$E_{1}(\theta)$ corresponds to Fig.~\ref{fig:magconfigs}(a) and $E_{2}(\theta)$
to Fig.~\ref{fig:magconfigs}(b). Note that the ring exchange enters as a term
in Eq.~\eqref{eq:dft_nc_rot1}(a) \footnote{The effect of the (square) ring
  exchange on the biquadratic interaction in the Fe-based superconductors has,
  to our knowledge, not been previously investigated. In
  Ref.~\onlinecite{Glasbrenner2014_PhysRevB_Firstprinciples}, rotations between
  the degenerate $\mathbf{q} = (0, \pi)$ and $\mathbf{q} = (\pi, 0)$ patterns
  are modeled as $E \sim K_{1} \sin^{2}(\theta)$. If the ring exchange is
  included, the model becomes $E \sim (K_{1} - 2 R) \sin^{2}(\theta)$. If
  $R > 0$, which for example is the case in FeTe, then not including ring
  exchange in single stripe rotations leads to an \textit{underestimation} of
  $K_{1}$. The opposite is true for the double stripe fluctuations, where not
  including it leads to an \textit{overestimation} of $K_{1}$}, which we take
from our collinear fits.

We used the following parameters in the calculations obtained using
\textsc{elk}. For the k-mesh, we used a $14 \times 14 \times 8$ k-mesh for the
ferromagnetic and checkerboard unit cells, a $12 \times 8 \times 8$ k-mesh for
the double stripe cell (also used for noncollinear rotation in
Fig.~\ref{fig:magconfigs}(a)), a $12 \times 8 \times 6$ k-mesh for the parallel
stripes unit cell, a $14 \times 14 \times 4$ k-mesh for the antiferromagnetic
layers unit cell, and a $8 \times 8 \times 8$ k-mesh for the plaquette unit
cell (also used for noncollinear rotation in Fig.~\ref{fig:magconfigs}(b)). The
number of empty states was set to 6 states/atom/spin. In addition, because of
how \textsc{elk} evaluates the exchange-correlation potential, convergence of
the $\theta= 0$ and $\theta= 180^{\circ}$ configurations in
Fig.~\ref{fig:magconfigs}(b) (which are supposed to be degenerate) required
setting the angular momentum cutoff for the APW functions (parameter lmaxapw)
and the muffin-tin density and potential (parameter lmaxvr) to 10, and also
reducing the fracinr parameter to 0.001 \footnote{Kay Dewhurst, the developer
  for \textsc{elk}, explained in a personal communication that the reason the
  symmetry between $\theta= 0$ and $\theta= 180^{\circ}$ configurations is
  slightly broken is because the exchange-correlation potential (and density in
  the case of Elk) is evaluated on a grid in real space, i.e.~not spherical
  harmonics. There is no way to evenly distribute $N$ points on the sphere
  while maintaining the symmetry. To limit this effect, the number of real
  space points has to be large, which can be achieved by scaling up the
  parameters lmaxapw and lmaxvr and scaling down fracinr.}.

For our fittings to Eq.~\eqref{J1_J2_J3}, as mentioned in the main text, the
oxygen sites in BaTi$_{2}$As$_{2}$O lead to an anisotropy in the Ti-Ti NNN
couplings, which in principle splits the second-neighbor Heisenberg exchange
parameter ($J_{2} \to J_{2O}, J_{2v}$), biquadratic exchange parameter
(($K_{2} \to K_{2O}, K_{2v}$)), and ring exchange parameter
(($R \to R_{O}, R_{v}$)). For consistency, we define a set of averaged exchange
parameters to use when fitting: $2 J_{2} = J_{2v} + J_{2O}$ for NNN Heisenberg
exchange, $2 K_{2} = K_{2O} + K_{2v}$ for NNN biquadratic exchange, and
$2 R = R_{O} + R_{v}$ for NNN ring exchange.

For our \textsc{vasp} structural relaxations, we used a plane wave energy
cutoff of 600 eV. We also used the same k-meshes for the different supercell
geometries as was used in the \textsc{elk} calculations.

\subsection{Total energy calculations and fitted exchange parameters}
\label{sec:total-energy-calc}

\begin{figure*}
\centering
\includegraphics[width=0.98\textwidth]{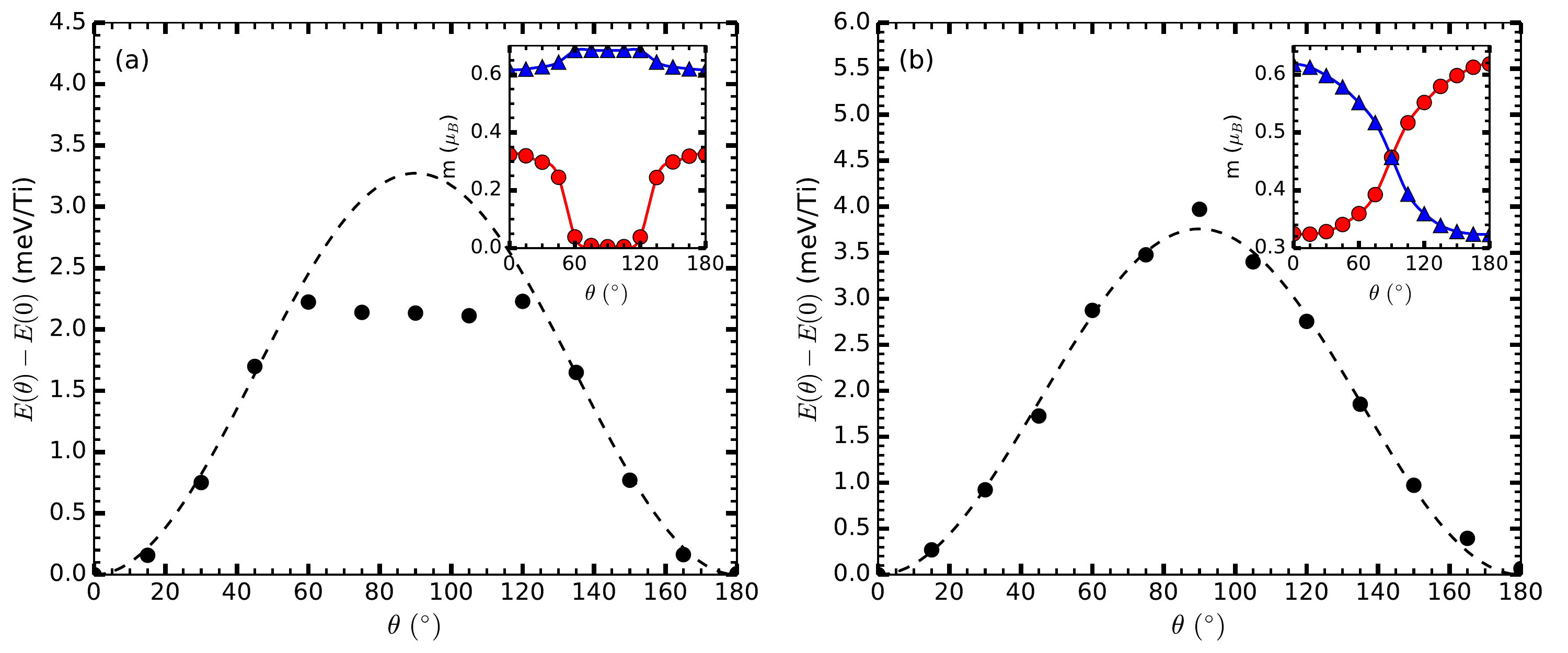}
\caption{Noncollinear energies as a function of the rotation angle
  $\theta$. The dashed lines are the fits to the model. The insets in each
  panel show the inequivalent Ti local moments as a function of $\theta $. The
  red circles refer to NNN FM O-bridging Ti sites and the blue triangles to NNN
  FM vacancy-bridging Ti sites when $\theta = 0$. (a) Noncollinear energies for
  the rotations analogous to fluctuations of the $\psi_{\pm}$ order parameter
  (see Fig.~\ref{fig:magconfigs}(a)). (b) Noncollinear energies for the
  rotations analogous to fluctuations of the $\varphi$ order parameter (see
  Fig.~\ref{fig:magconfigs}(b))}
\label{fig:biquadenergy}
\end{figure*}

Table \ref{tab:collinear_energy} contains the full summary of our total energy
calculations and the Ti local moment amplitudes of the different magnetic
patterns. We find that our results are consistent with the trends reported in
Ref.~\onlinecite{Wang2013_JApplPhys_electronic}, where increasing $U$ lowers
the energy of each configuration and increases the amplitude of the local
moments. The local moments themselves are soft and can vary by more than a
factor of 2 between magnetic patterns. Our calculations also capture the energy
difference that arises due to the anisotropy in the $J_{2}$ parameter, which
depends on whether the NNN ferromagnetic bonds bridge either an oxygen site or
a vacancy. Overall NNN ferromagnetic bonds are energetically preferred.

The results of our noncollinear energy calculations are shown in
Fig.~\ref{fig:biquadenergy}, which we fit to
Eqs.~\eqref{eq:dft_nc_rot1}--\eqref{eq:dft_nc_rot2} to obtain the biquadratic
parameters. In Fig.~\ref{fig:biquadenergy}(a) the $M_{2}$ Ti moments collapse
when $60^{\circ} \leq \theta \leq 120^{\circ}$; for simplicity we fit to
Eq.~\eqref{eq:dft_nc_rot1} using only the energy calculations obtained for
$\theta$ outside this range. As a side note, the rotations in
Fig.~\ref{fig:magconfigs} are analogous to the fluctuations discussed in
Sections \ref{sec:model-intro} and \ref{sec:model_model}, with
Fig.~\ref{fig:magconfigs}(a) being similar to fluctuations between the $(++-+)$
and $(-+++)$ states in Fig.~\ref{fig:table_four} that are frozen out when
$T < T_{\psi}$, and Fig.~\ref{fig:magconfigs}(b) being similar to fluctuations
between the $(++-+)$ and $(+++-)$ states in Fig.~\ref{fig:table_four} that are
frozen out when $T < T_{\varphi}$.

\begin{table*}
  \centering
  \begin{tabular}{c c c c c c c c c}
    \toprule
    Functional & $U$ & $J_{1}$ & $J_{2}$ & $J_{3}$ & $J_{\perp}$ & $R$ & $K_{1}$ & $K_{2}$ \\
             & (eV) & \multicolumn{7}{c}{(meV)} \\
    \midrule
    LDA+U & 2.5 & 0.076 & -1.04 & 1.59 &  0.32 & 0.38 & & \\
    LDA+U & 3.5 & 0.89  & -2.83 & 2.79 & -0.26 & 1.00 & -0.37 & 2.06 \\
    GGA+U & 2.5 & 1.66  & -2.41 & 1.89 &  0.52 & 0.92 & & \\
    \bottomrule
  \end{tabular}
  \caption{The calculated exchange parameters for BaTi$_{2}$Sb$_{2}$O.}
  \label{tab:btso_exchange_params}
\end{table*}

The fitted exchange parameters are summarized in Table
\ref{tab:btso_exchange_params}. For completeness, we included the interplanar
coupling $J_{\perp}$ that was neglected in the model treatment. The sign of the
Heisenberg and ring exchange parameters are consistent across functionals and
values of $U$ with the exception of $J_{\perp}$, which is slightly
ferromagnetic in LSDA$+U = 3.5 \text{ eV}$, but antiferromagnetic otherwise.

\subsection{Structural relaxation data}
\label{sec:appendix_B}

\begin{table*}[!htbp]
\centering
\begin{tabular}{cccccccc}
  \toprule Functional & $U - J$ & Distortion & Energy & $Q(\text{Ti}_{1})$ & $Q(\text{Ti}_{2})$ & $\theta_{\text{Ti}}$ & $\eta$ \\ 
                      & (eV) &  & (eV/Ti) & \multicolumn{2}{c}{(elec.)} & (deg) & \% \\ 
  \midrule GGA & 0.0 & None & -19.59 & 10.2 & 10.2 & 0.0 & 0.0 \\ 
  GGA & 0.0 & Rotation & -19.59 & 10.2 & 10.2 & 3.5 & 0.028 \\ 
  GGA & 0.0 & Orthorhombic & -19.59 & 10.2 & 10.2 & 0.0 & 0.40 \\ 
  LSDA+U & 3.0 & None & -19.48 & 10.1 & 10.2 & 0.0 & 0.0 \\ 
  LSDA+U & 3.0 & Rotation & -19.48 & 10.2 & 10.2 & 2.19 & 0.08 \\ 
  LSDA+U & 3.0 & Orthorhombic & -19.48 & 10.1 & 10.2 & 0.0 & 0.51 \\ 
  LSDA+U & 3.0 & Double stripe & -19.49 & 10.2 & 10.1 & 0.0 & 1.4 \\ 
  \bottomrule
\end{tabular}
\caption{A summary of the results of structural relaxations of
  BaTi$_{2}$Sb$_{2}$O for different exchange-correlation functionals and kinds
  of distortions. The DFT+U calculations used the rotationally invariant
  approach with a single parameter $U - J$
  \cite{Dudarev1988_PhysRevB_Electronenergyloss}. The $Q(\text{Ti}_{1})$ and
  $Q(\text{Ti}_{2})$ columns report the calculated charge on the inequivalent
  Ti sites, the $\theta_{\text{Ti}}$ column reports how many degrees the
  oxygen-centered Ti plaquettes are rotated in each distortion (if at all), and
  the final column calculates $\eta = 2 \cdot \frac{a-b}{a+b} \cdot 100\%$,
  which quantifies the degree of the orthorhombic distortion.}
\label{tab:btso_vasp_relaxation}
\end{table*}

The structural relaxations we computed using \textsc{vasp} are summarized in
Table \ref{tab:btso_vasp_relaxation}. For all functionals we first performed a
baseline relaxation calculation where we enforced the high temperature
structure with $P4/mmm$ symmetry. We then considered up to three kinds of
distortions. The nonmagnetic distortions are the ``rotation'' distortion, which
refers to rotations of Ti plaquettes about the oxygen sites by an angle
$\theta_{\text{Ti}}$, and the ``orthorhombic'' distortion, which is a spitting
of the $a$ and $b$ lattice parameters quantified with the parameter
$\zeta = 2 \cdot \frac{a-b}{a+b} \cdot 100\%$. The ``double stripe'' distortion,
on the other hand, is obtained by performing a structural relaxation for the
magnetic double stripe pattern. We then compared the energies, the calculated
charges on the two inequivalent Ti sites, and the distortion parameters
$\theta_{\text{Ti}}$ and $\zeta$.

We found that for the vanilla DFT calculations with the GGA functional, the
plaquette rotation distortion is lowest in energy, with E(rotation) - E(none) =
-5.8 meV/Ti compared with E(orthorhombic) - E(none) = 0.03/Ti meV.  The
undistorted structures feature a charge imbalance on the inequivalent Ti sites,
while the distorted sites do not. The rotated plaquettes structure also has a
minor orthorhombic distortion of 0.03\%, which is negligible.

For the LSDA+U (U - J = 3.0 eV) calculations, the double stripe distortion is
clearly the lowest in energy, with E(ds) - E(none) = -18.4 meV/Ti compared with
E(rotation) - E(none) = -0.88 meV/Ti and E(orthorhombic) - E(none) = -0.079
meV/Ti. The nonmagnetic distortions do not provide much of an energy gain,
particularly when compared with the relaxed magnetic state. As in the case of
vanilla GGA, the nonmagnetic distortions remove the charge imbalance between
$M_{1}$ and $M_{2}$ found in the high temperature structure. In contrast, the
charge imbalance still persists after relaxing the double stripe pattern.

In terms of symmetry breaking, only the relaxed magnetic calculations break
rotational, reflection, and translational symmetry, induce a orthorhombic
lattice distortion, and preserve a charge imbalance between the inequivalent Ti
sites. The nonmagnetic distortions may or may not break the right symmetries
compared with experiment, and after relaxation the charge imbalance disappears.

\subsection{BaTi$_{2}$As$_{2}$O}
\label{sec:appendix_C}

\begin{table}[!htbp]
\centering
\begin{tabular}{cccc}
  \toprule Config & E - E(NM) & $M_{1}$ & $M_{2}$ \\ 
                  & (meV/Ti) & \multicolumn{2}{c}{$(\mu_{B})$} \\ 
  \midrule FM & -0.8245 & 0.3116 & 0.3116 \\ 
  DS (O-FM only) & -4.562 & 0.3226 & 0.0000 \\ 
  DS (v-FM only) & -9.787 & 0.0000 & 0.4673 \\ 
  Parallel & -3.354 & 0.1987 & 0.1987 \\ 
  Plaquette (O-centered) & -4.843 & 0.2359 & 0.2359 \\ 
  Plaquette (v-centered) & -9.246 & 0.3260 & 0.3260 \\ 
  AFM Layers & -3.257 & 0.3242 & 0.3242 \\ 
  \bottomrule &  &  & 
\end{tabular}
\caption{The magnetic energies for the collinear magnetic patterns that are
  stable in BaTi$_{2}$As$_{2}$O for LSDA+U with $U = 3.5 \text{ eV}$ and
  $J = 0.5 \text{ eV}$. Energies are referenced against the nonmagnetic (NM)
  state.  The inequivalent magnetic moments $M_{1}$ and $M_{2}$ in the double
  stripe and site-selective patterns are the same as those labeled in
  Fig.~\ref{fig:colpatterns}}
\label{tab:btao_collinear_energy}
\end{table}

\begin{figure}[!htbp]
\centering
\includegraphics[width=0.48\textwidth]{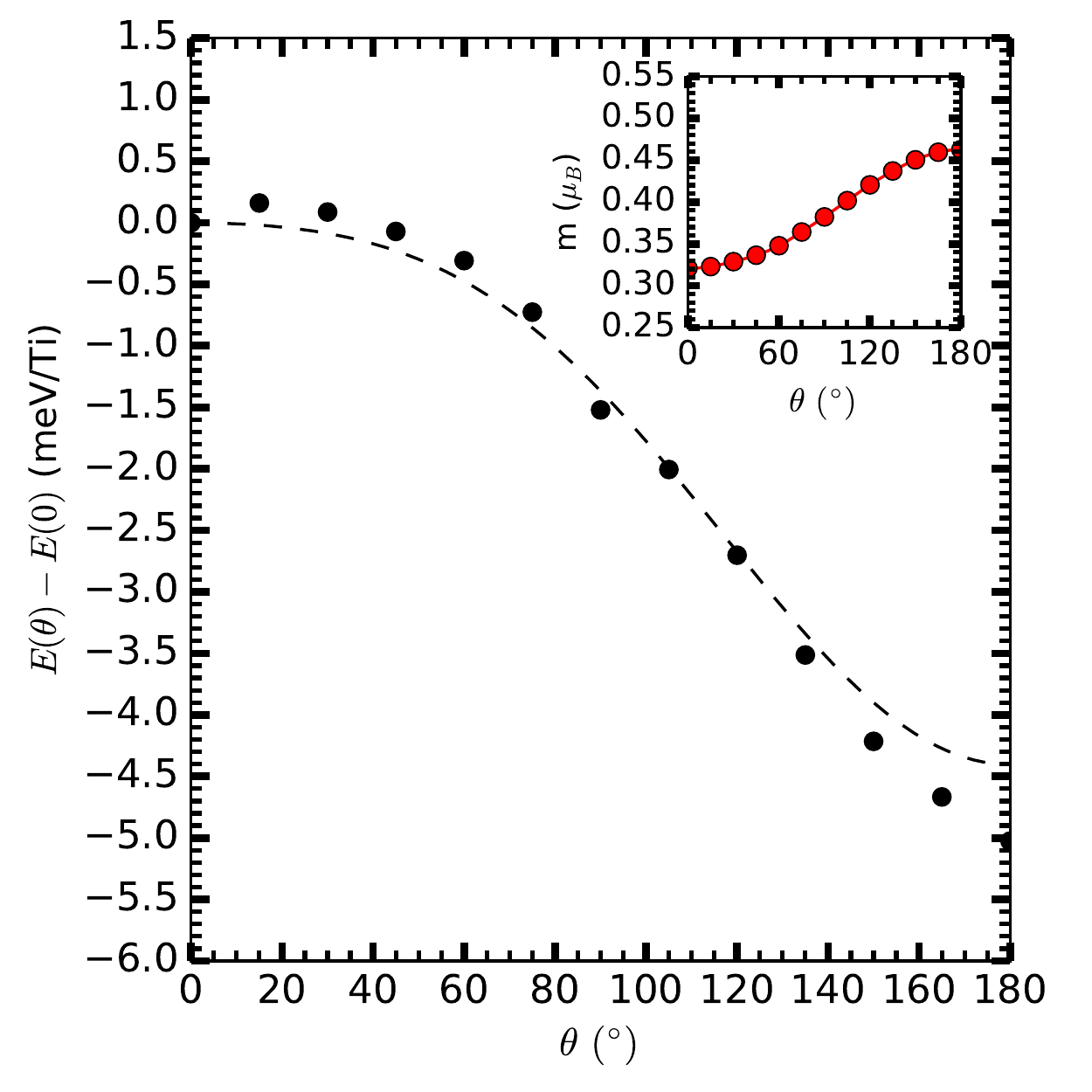}
\caption{Noncollinear energies for rotations connecting the two double stripe
  patterns where one of the local Ti moments has collapsed. $\theta = 0$
  corresponds to moments with oxygen-bridging FM NNN interactions and
  $\theta = 180^{\circ}$ to moments with vacancy-bridging FM NNN interactions.
  The dashed line is the fit to the model. The inset shows the Ti local moment
  as a function of $\theta$.}
\label{fig:btao_biquadenergy}
\end{figure}

We performed a set of DFT calculations for BaTi$_{2}$As$_{2}$O in order to
compare with the main BaTi$_{2}$Sb$_{2}$O results. These calculations used the
LSDA+U functional with $U = 3.5 \text{ eV}$ and $J = 0.5 \text{ eV}$. We used
the experimental crystal structure for these calculations
\cite{Wang2010_JPhysCondensMatter_Structure}, which has space group symmetry
$P4/mmm$, lattice parameters a = 4.04561 Åand c = 7.27228 Å, and the following
Wyckoff positions in fractional coordinates: Ba [1d] (0.5, 0.5, 0.5), Ti [2f]
(0.5, 0, 0), As [2g] (0, 0, 0.7560), and O [1c] (0.5, 0.5, 0). We used the same
k-meshes and parameters as was used for BaTi$_{2}$Sb$_{2}$O.

The results of the collinear calculations are summarized in Table
\ref{tab:btao_collinear_energy}. BaTi$_{2}$As$_{2}$O is less supportive of
magnetism compared to BaTi$_{2}$Sb$_{2}$O, as the full double stripe and
checkerboard patterns cannot be stabilized and the stable patterns yield less
of an energy gain compared to their BaTi$_{2}$Sb$_{2}$O counterparts. Because
of this, there are not enough stable collinear magnetic patterns that we can
use to fit to Eq.~\eqref{J1_J2_J3}. Trying to include the patterns with
nonmagnetic sites further complicates the model, as we would need to add
Stoner-like on-site terms to capture variations in the local moments. 

Even though we cannot resolve all the exchange parameters through a fit, we can
at least estimate the NNN biquadratic parameter. We do this by performing
noncollinear calculations with rotations that interpolate between the two kinds
of double stripe patterns where half the sites are nonmagnetic. The results of
these calculations are presented in Fig.~\ref{fig:btao_biquadenergy}. Applying
Eq.~\eqref{J1_J2_J3}, we obtain the following expression,
\begin{align}
  \label{eq:btso_noncol}
  E(\theta) - E(0) &= 2 K_{2} \sin^{2}\theta
                     + 2 \left( J_{2v} - J_{2O} \right) \sin^{2} \left( \frac{\theta}{2} \right)
\end{align}
In Eq.~\eqref{eq:btso_noncol} the anisotropic splitting of $J_{2}$ enters as a
\textit{difference} instead of a sum, so we can't use the average value $J_{2}$
here. However, we also note that the energy difference between the two
plaquette configurations is
$E(\text{Plaq}_v) - E(\text{Plaq}_O) = 2 \left( J_{2v} - J_{2O} \right)$, which
can be substituted in Eq.~\eqref{eq:btso_noncol} to allow us to resolve
$K_{2}$. We obtain $K_{2} = 0.418 \text{ meV}$ from this fit, but without
$J_{2v}$, $J_{2O}$ and $J_{3}$ available for comparison, it is unclear what
regime of the model in Section \ref{sec:model_model} BaTi$_{2}$As$_{2}$O is in.

\bibliography{btso_nematics}

\end{document}